\documentclass[11pt,reqno]{amsart}

\usepackage[T1]{fontenc}
\usepackage[utf8]{inputenc}
\usepackage{lmodern}
\usepackage{microtype}
\usepackage{mathtools,amssymb,amsthm}
\usepackage{enumitem}
\usepackage{xcolor}
\usepackage{calc}
\usepackage{natbib}
\IfFileExists{xurl.sty}{\usepackage{xurl}}{}
\usepackage{hyperref}
\usepackage[nameinlink,noabbrev]{cleveref}
\hypersetup{
  colorlinks=true,
  linkcolor=blue,
  citecolor=blue,
  urlcolor=blue,
  pdftitle={Axiomatic Shared-Medium Coordination for Stigmergic
Systems},
  pdfauthor={Fernando Paredes García}
}

\providecommand{\tightlist}{%
  \setlength{\itemsep}{0pt}\setlength{\parskip}{0pt}}

\theoremstyle{plain}
\newtheorem*{namedplaininner}{\namedplaincaption}
\newcommand{\namedplaincaption}{Result}
\newenvironment{namedplain}[1]{\renewcommand{\namedplaincaption}{#1}\begin{namedplaininner}}{\end{namedplaininner}}

\theoremstyle{definition}
\newtheorem*{nameddefinitioninner}{\nameddefinitioncaption}
\newcommand{\nameddefinitioncaption}{Definition}

\theoremstyle{remark}
\newtheorem*{namedremarkinner}{\namedremarkcaption}
\newcommand{\namedremarkcaption}{Remark}
\newenvironment{namedremark}[1]{\renewcommand{\namedremarkcaption}{#1}\begin{namedremarkinner}}{\end{namedremarkinner}}

\title{Axiomatic Shared-Medium Coordination for Stigmergic Systems}
\author{Fernando Paredes García}
\address{Independent Researcher}
\email{fernando@develcuy.com}

\begin{document}
\begin{abstract}
Stigmergic coordination has many medium-specific formalizations. This
paper formulates a medium-agnostic comparison layer whose comparison
object is the abstract enabled-response signature. For metadata
refinements, we prove that a coarse quotient is response-adequate
exactly when each quotient fiber is response-aligned. When that fails,
there is a canonical coarsest response-adequate repair, and every
response-adequate repair must retain at least the corresponding
fiberwise number of response classes. On the dynamic side,
quotient-compatible matched writes preserve one-step behavior,
freshness-conditional actions yield a one-step non-liftability
obstruction, and the matched one-step result lifts to finite serial
traces. These results are instantiated for a tuple-space medium and a
timestamped virtual-stigmergic dataspace. The result is a
medium-agnostic comparison framework with one explicit positive/negative
theorem cycle; richer pathwise and concurrent semantics remain outside
scope.
\end{abstract}
\maketitle

\hypertarget{introduction}{%
\section{1. Introduction}\label{introduction}}

Stigmergic coordination is older and broader than any single engineering
substrate. The term originates in the biological literature on indirect
coordination through environmental traces \citep{grasse1959}, later
survey work established stigmergy as a family of indirect coordination
mechanisms rather than an insect-specific curiosity
\citep{theraulaz1999history}, and more recent conceptual accounts
decomposed stigmergy into agents, actions, traces, media, and
coordination effects
\citep{heylighen2016stigmergy1, heylighen2016stigmergy2}. In parallel,
several computational lines gave explicit semantics to engineered shared
media, including shared dataspaces, virtual stigmergy, linked-data
coordination environments, and control-oriented mathematical models for
trace-mediated swarm behavior
\citep{gelernter1985linda, carriero1989linda, denicola2020virtual, linkeddata2021, stigld2022, boldini2024control}.
The basic phenomenon is therefore not missing from the literature, nor
are formal models absent in every domain.

What remains incomplete is narrower and more technical. The literature
already contains conceptual universality claims, trace semantics,
expressiveness comparisons within shared-dataspace coordination families
\citep{heylighen2016stigmergy1, heylighen2016stigmergy2, tummolini2007trace, dipple2014general, brogi2003expressiveness},
and medium-specific operational models
\citep{denicola2020virtual, distefano2023verification, linkeddata2021, stigld2022}.
We are not aware of a theorem-bearing comparison treatment centered on
enabled-response signatures, metadata-refinement quotients, canonical
repairs, and repair lower bounds across heterogeneous shared media. The
present work addresses that narrower gap.

The contribution is therefore not a claim to invent stigmergy, to give
the first mathematical treatment of the topic, or to supersede the
existing semantic and control literatures. Rather, we develop an
axiomatic framework for shared-medium coordination together with one
explicit comparison theorem cycle, and instantiate it for a
shared-dataspace or tuple-space medium paired with a timestamped
virtual-stigmergic dataspace. The main results include a quotient
adequacy criterion, a canonical coarsest repair construction with
fiberwise lower bounds, one-step update preservation for matched write
actions that commute with the quotient, a one-step non-liftability
obstruction for freshness-conditional refined actions, and the finite
serial closure of the matched one-step result under an explicit
matched-action relation.

The core tuple is axiomatic and medium-centered by design. Readers from
theoretical computer science may recognize a family resemblance to
general reactive-system formalisms such as input/output automata,
reactive modules, and TLA-style state-machine specifications
\citep{lynchtuttle1989ioa, alur1999reactive, lamport2002specifying}.
That comparison is only orienting context, not the primary identity
here. The distinctive aim is to treat medium-mediated observation and
enabled response as the central comparison object, and to make
representation between heterogeneous media a theorem-bearing question,
not only a semantic analogy.

The manuscript is deliberately not organized around ledger-state
stigmergy, even though ledger systems remain a useful downstream
specialization
\citep{gurcan2022pow, capaccioli2020blockchain, dounas2022stigmergic}.
The scope is likewise bounded: we develop statewise, one-step, and
finite serial matched-trace comparison results, not a full
branching/pathwise equivalence theory, scheduler-robust liveness theory,
or finished control synthesis. Those stronger directions remain outside
the present treatment.

The rest of the paper proceeds as follows. Section 2 positions the
manuscript against the main prior-art clusters. Section 3 defines the
abstract system class and its primitive objects. Section 4 introduces
admissibility classes. Section 5 records a descriptive operator taxonomy
used later for interpretation. Section 6 formulates the
response-comparison layer. Section 7 develops structural results
together with metadata-refinement theorems, one matched-action
obstruction, repair lower bounds, and one finite serial closure result.
Section 8 works out the tuple-space and virtual-stigmergy pair
explicitly. Section 9 records the present limits and Section 10
concludes.

\hypertarget{related-work-and-positioning}{%
\section{2. Related Work and
Positioning}\label{related-work-and-positioning}}

\hypertarget{historical-and-conceptual-foundations}{%
\subsection{2.1 Historical and conceptual
foundations}\label{historical-and-conceptual-foundations}}

Any foundations paper on stigmergy must start from the fact that the
concept is already well established. The term enters through the
biological literature \citep{grasse1959}, later surveys consolidate
stigmergy as a family of indirect coordination mechanisms
\citep{theraulaz1999history}, and universal conceptual accounts
decompose the phenomenon into agents, actions, traces, media, and
coordination effects
\citep{heylighen2016stigmergy1, heylighen2016stigmergy2}. We do not
compete on universal conceptual discovery. The intended addition is
narrower: a systems language in which admissibility and representation
questions become mathematically expressible across media.

\hypertarget{trace-semantics-and-general-theory-attempts}{%
\subsection{2.2 Trace semantics and general-theory
attempts}\label{trace-semantics-and-general-theory-attempts}}

Trace-signal work makes clear that traces are meaningful environmental
signals rather than mere residues \citep{tummolini2007trace}. Related
``general theory'' work already pushes toward cross-domain vocabulary
unification \citep{dipple2014general}. These contributions block any
sweeping rhetoric about a first general theory. The distinction pursued
here is instead structural: we seek a state-transition comparison layer
with explicit update operators, response-sufficient features, and
representation theorems.

\hypertarget{engineered-shared-media}{%
\subsection{2.3 Engineered shared media}\label{engineered-shared-media}}

Engineered digital media already have serious formal treatment. The
shared-dataspace and coordination-language lineage shows that
environment-mediated coordination is not new in computer science
\citep{gelernter1985linda, carriero1989linda, omicini2004artifacts, weyns2007environment}.
Shared-dataspace work also includes expressiveness comparisons across
Linda-style coordination families \citep{brogi2003expressiveness}.
Virtual-stigmergy work then makes the connection explicit in distributed
robotic and multi-agent settings
\citep{pinciroli2016virtual, denicola2020virtual}, while
verification-oriented papers equip such media with timestamped
propagation semantics and operational models
\citep{distefano2023verification}. Linked-data work and the stigLD line
show that shared stigmergic media can also be formalized with
graph-oriented domain semantics \citep{linkeddata2021, stigld2022}.
These are the nearest formal neighbors of the present project. They also
set a tight novelty boundary: a medium-agnostic theory cannot restate
one digital substrate or compare only variants inside the
shared-dataspace family; it must compare multiple substrates inside one
system class.

\hypertarget{mathematical-modelling-and-relation-to-shared-state-abstraction}{%
\subsection{2.4 Mathematical modelling and relation to shared-state
abstraction}\label{mathematical-modelling-and-relation-to-shared-state-abstraction}}

Recent mathematical work on stigmergy in swarm and control settings
demonstrates that stigmergy is already a mathematical topic in its own
right \citep{boldini2024control}. For orientation, the tuple used here
can be compared with general reactive-system formalisms: input/output
automata model concurrent components through state and action interfaces
\citep{lynchtuttle1989ioa}, reactive modules provide a
guarded-transition language for modular reactive systems
\citep{alur1999reactive}, and TLA+ exemplifies state-machine
specification of concurrent systems at the action level
\citep{lamport2002specifying}. These are adjacent reference points, not
the governing template. The narrower claim pursued here is that, once
the shared state is interpreted as a coordination medium,
medium-mediated observation and enabled response become the comparison
object, and cross-medium representation becomes a theorem-bearing
problem.

\hypertarget{positioning-statement}{%
\subsection{2.5 Positioning statement}\label{positioning-statement}}

The safe and substantive claim is therefore the following:

\begin{quote}
We develop an axiomatic comparison layer for stigmergic coordination
over shared media, formulated in explicit update-and-observation terms
and intended to support structural comparison, quotient construction,
and realizability questions across multiple realizations of indirect
coordination.
\end{quote}

That claim leaves historical foundations, taxonomic universality,
semantic verification, domain-specific modelling, and ledger-specific
application work in their proper places. It also defines the standard
against which the rest of the manuscript must be judged: not by breadth
of examples alone, but by the mathematical substance of the comparison
results.

\hypertarget{mathematical-setting}{%
\section{3. Mathematical Setting}\label{mathematical-setting}}

We model stigmergic coordination over a shared medium whose visible
state changes under actions and whose resulting traces influence later
actions. The decomposition is compatible with the canonical
agent-action-medium-trace picture of the conceptual literature, but it
is recast here in an explicit axiomatic update-and-observation language
\citep{heylighen2016stigmergy1}.

\hypertarget{primitive-objects}{%
\subsection{3.1 Primitive objects}\label{primitive-objects}}

Let:

\begin{itemize}
\tightlist
\item
  \(\mathcal{M}\) be the set of admissible medium states.
\item
  \(\mathcal{A}\) be the action alphabet.
\item
  \(\mathcal{I}\) be the set of agents.
\item
  \(\mathcal{V}_i\) be the observation space available to agent
  \(i \in \mathcal{I}\).
\end{itemize}

We write the medium state at step \(t\) as \(m_t \in \mathcal{M}\).

\hypertarget{core-tuple}{%
\subsection{3.2 Core tuple}\label{core-tuple}}

A shared-medium stigmergic system is specified by

\[
\mathfrak{S} = \langle \mathcal{M}, \mathcal{A}, \mathcal{I}, U, \{O_i\}_{i \in \mathcal{I}}, \{P_i\}_{i \in \mathcal{I}}, \{R_i\}_{i \in \mathcal{I}} \rangle.
\]

Here:

\begin{itemize}
\tightlist
\item
  \(U : \operatorname{Adm} \subseteq \mathcal{M} \times \mathcal{A} \to \mathcal{M}\)
  is the update operator on admissible state-action pairs, where
  \(\operatorname{Adm}\) is part of the specified system data.
\item
  \(O_i : \mathcal{M} \to \mathcal{V}_i\) is the observation operator
  for agent \(i\).
\item
  \(P_i : \mathcal{V}_i \to \{0,1\}\) is the activation predicate for
  agent \(i\).
\item
  \(R_i : \mathcal{V}_i \to \mathcal{A} \cup \{\bot\}\) is the response
  map for agent \(i\), where \(\bot\) denotes inaction.
\end{itemize}

The induced medium-to-action map for agent \(i\) is

\[
\Gamma_i = R_i \circ O_i : \mathcal{M} \to \mathcal{A} \cup \{\bot\}.
\]

Enabled behavior can be expressed by the set-valued map

\[
E_{\mathfrak{S}}(m) = \{(i,a) \in \mathcal{I} \times \mathcal{A} : P_i(O_i(m)) = 1 \text{ and } R_i(O_i(m)) = a\}.
\]

This object is useful later because it records the response structure
induced by the medium state without presupposing any pairwise
communication graph. Because each \(R_i\) is a function, we work with
deterministic one-response policies at the comparison layer; set-valued
or randomized response rules lie outside the current scope. The response
structure is observational rather than executable by itself: later
dynamic statements add the separate requirement that the selected action
lie in the admissibility domain of \(U\).

Standing convention. Throughout, \(R_i(v) = \bot\) whenever
\(P_i(v) = 0\). That is, the response map returns inaction on any
observation for which the agent is not enabled. This ensures that
\(\Gamma_i(m) = \bot\) whenever \(\beta_i(m) = 0\), so the response
label carries no information beyond the activation predicate in disabled
states.

\hypertarget{traces}{%
\subsection{3.3 Traces}\label{traces}}

A trace is any medium-visible successor feature capable of affecting a
later observation. We do not fix one privileged encoding of traces; at
the abstract level it is enough to work with the before-after pair
generated by an admissible action:

\[
\tau(m,a) = (m, U(m,a)),
\]

or, when convenient, by the successor-difference it induces. This
deliberately leaves open whether the underlying encoding is geometric,
chemical, symbolic, graph-structured, or ledger-based.

\hypertarget{coordination-loop}{%
\subsection{3.4 Coordination loop}\label{coordination-loop}}

The abstract stigmergic loop is:

\begin{enumerate}
\def\labelenumi{\arabic{enumi}.}
\tightlist
\item
  some agent \(j\) executes an admissible action \(a_j\) at state
  \(m_t\);
\item
  the medium updates to \(m_{t+1} = U(m_t, a_j)\);
\item
  another agent \(i\) observes \(O_i(m_{t+1})\);
\item
  if \(P_i(O_i(m_{t+1})) = 1\), then agent \(i\) selects
  \(R_i(O_i(m_{t+1}))\);
\item
  the selected action, if non-null and admissible, generates the next
  medium update.
\end{enumerate}

No direct message channel between agents is required for this loop.
Direct messaging may coexist with it in particular systems, but the
present theory tracks only the coordination that is mediated through the
shared medium.

\hypertarget{admissibility-classes}{%
\section{4. Admissibility Classes}\label{admissibility-classes}}

The core tuple becomes mathematically informative only after
admissibility conditions are fixed. The conditions below are intended as
the minimal class assumptions that survive abstraction across media.

\hypertarget{base-admissibility}{%
\subsection{4.1 Base admissibility}\label{base-admissibility}}

We call \(\mathfrak{S}\) an admissible stigmergic system if:

\begin{enumerate}
\def\labelenumi{\arabic{enumi}.}
\tightlist
\item
  \(\operatorname{Adm}\) is a specified admissible subset of
  \(\mathcal{M} \times \mathcal{A}\).
\item
  \(U\) is closed on admissible pairs: if
  \((m,a) \in \operatorname{Adm}\), then \(U(m,a) \in \mathcal{M}\).
\item
  each observation operator \(O_i\) is total on \(\mathcal{M}\).
\item
  each activation predicate \(P_i\) depends only on \(O_i(m)\).
\item
  each response map \(R_i\) depends only on \(O_i(m)\).
\end{enumerate}

Conditions (4) and (5) encode the locality that makes the system
stigmergic rather than centrally coordinated. The theory does not
require that agents have identical observations, only that each agent's
activation and response be functions of the medium-visible information
available to that agent. The framework treats \(\operatorname{Adm}\) as
system-specific data; no canonical construction of it from
\(\mathcal{M} \times \mathcal{A}\) is assumed. Accordingly, ``enabled
response'' in the comparison layer means enabled by the
observation-response operators; executable one-step dynamics impose the
additional requirement that the selected action also lie in
\(\operatorname{Adm}\).

\hypertarget{persistence-and-decay}{%
\subsection{4.2 Persistence and decay}\label{persistence-and-decay}}

The distinction between persistent and decaying traces cuts across
natural and engineered media
\citep{theraulaz1999history, heylighen2016stigmergy2}.

Definition 4.1. A persistent-trace system is an admissible system in
which traces remain encoded in the successor state until altered by
later admissible updates.

Definition 4.2. A decaying-trace system is an admissible system equipped
with a decay operator \(D : \mathcal{M} \to \mathcal{M}\) such that the
inter-step evolution is

\[
m_{t+1} = D(U(m_t, a_t)).
\]

The operator \(D\) separates agent-induced modification from
medium-induced weakening, evaporation, expiry, or forgetting. Biological
pheromone systems typically require nontrivial decay; persistent
digital-media models are often captured by the special case
\(D = \mathrm{id}\) when expiry or forgetting is not built into the
medium semantics.

\hypertarget{exclusive-resource-systems}{%
\subsection{4.3 Exclusive-resource
systems}\label{exclusive-resource-systems}}

Many engineered stigmergic systems manage resources that should not be
claimed by multiple agents simultaneously. Let \(Q\) be a family of
resources. For each \(q \in Q\), let:

\begin{itemize}
\tightlist
\item
  \(s_q : \mathcal{M} \to \Sigma_q\) be a status map,
\item
  \(c_q : \mathcal{M} \to \mathcal{I} \cup \{\bot\}\) be an owner map.
\end{itemize}

Definition 4.3. An admissible system is an exclusive-resource system if
there exists a distinguished status value \(\mathrm{CLAIMED}\) such that

\[
\forall q \in Q,\ \forall m \in \mathcal{M},\quad |\{i \in \mathcal{I} : c_q(m) = i \wedge s_q(m) = \mathrm{CLAIMED}\}| \leq 1.
\]

This single-occupancy condition abstracts claims, locks, reservations,
winning bids, and similar mutual-exclusion patterns.

\hypertarget{operator-structure}{%
\section{5. Operator Structure}\label{operator-structure}}

The recurring operator shapes below are descriptive rather than
theorem-bearing. They summarize update patterns that reappear across the
realizations and help interpret the later worked comparison without
carrying independent theorem weight.

\hypertarget{descriptive-operator-families}{%
\subsection{5.1 Descriptive operator
families}\label{descriptive-operator-families}}

We use four recurring labels:

\begin{itemize}
\tightlist
\item
  flag updates change finite-state status variables such as open,
  claimed, done, or expired;
\item
  signal updates append or publish traces intended for later
  observation;
\item
  threshold updates matter through order relations, scores, counts, or
  boundary crossings visible in the medium;
\item
  sequencing updates make admissibility depend on a protocol phase
  encoded in the medium.
\end{itemize}

These families are not mutually exclusive ontological categories. A
concrete system may combine them.

\hypertarget{observation-response-summaries}{%
\subsection{5.2 Observation-response
summaries}\label{observation-response-summaries}}

The pair \((O_i, R_i)\) induces the medium-to-action map
\(\Gamma_i = R_i \circ O_i\). This lets us treat agent behavior as a
family of response operators over the medium.

Threshold-driven behavior often factors as

\[
\Gamma_i(m) = \widehat{R}_i(\phi_i(O_i(m))),
\]

where \(\phi_i\) is a scalar-valued or partially ordered summary of the
observation. Such factorization matters because it exposes the
comparison object that may be preserved across otherwise different
media.

\hypertarget{role-of-the-taxonomy}{%
\subsection{5.3 Role of the taxonomy}\label{role-of-the-taxonomy}}

The operator families matter here for interpretation, not for proof
burden. They explain why seemingly different media can support analogous
coordination regimes, help organize the realizations in Section 8, and
keep medium-specific examples subordinate to the abstract theory. No
theorem below depends on a formal decision procedure for operator-family
membership.

\hypertarget{composition-and-equivalence}{%
\section{6. Composition and
Equivalence}\label{composition-and-equivalence}}

\hypertarget{product-composition}{%
\subsection{6.1 Product composition}\label{product-composition}}

Suppose

\[
\mathfrak{S}_k = \langle \mathcal{M}_k, \mathcal{A}_k, \mathcal{I}_k, U_k, O^k, P^k, R^k \rangle
\]

for \(k = 1, \dots, n\).

Define the product medium

\[
\mathcal{M} = \mathcal{M}_1 \times \cdots \times \mathcal{M}_n.
\]

A coupled system is then specified by:

\begin{itemize}
\tightlist
\item
  a joint observation operator \(O_i : \mathcal{M} \to \mathcal{V}_i\),
\item
  a joint activation predicate \(P_i(O_i(m))\),
\item
  a joint response map \(R_i(O_i(m))\),
\item
  an update rule \(U\) whose action may affect one or several
  coordinates.
\end{itemize}

This is the abstract form of cross-substrate or cross-protocol
stigmergy. Composition changes the topology of observable dependence
without changing the medium-trace-response architecture.

\hypertarget{a-comparison-ready-equivalence-notion}{%
\subsection{6.2 A comparison-ready equivalence
notion}\label{a-comparison-ready-equivalence-notion}}

The full theory of equivalence between media is left for later work, but
one comparison-ready notion is already useful.

Definition 6.1. Let

\[
\begin{aligned}
\mathfrak{S} &= \langle \mathcal{M}, \mathcal{A}, \mathcal{I}, U, \{O_i\}, \{P_i\}, \{R_i\} \rangle, \\
\mathfrak{S}' &= \langle \mathcal{M}', \mathcal{A}', \mathcal{I}, U', \{O'_i\}, \{P'_i\}, \{R'_i\} \rangle.
\end{aligned}
\]

be admissible stigmergic systems over a common agent set
\(\mathcal{I}\), with admissibility domains

\[
U : \operatorname{Adm} \subseteq \mathcal{M} \times \mathcal{A} \to \mathcal{M},
\qquad
U' : \operatorname{Adm}' \subseteq \mathcal{M}' \times \mathcal{A}' \to \mathcal{M}'.
\]

Let

\[
\alpha : \mathcal{A} \to \mathcal{A}^{\ast}, \qquad \alpha' : \mathcal{A}' \to \mathcal{A}^{\ast}
\]

be abstraction maps into a common comparison alphabet
\(\mathcal{A}^{\ast}\). Here \(\mathcal{A}^{\ast}\) denotes a fixed
abstract comparison alphabet shared by the comparison layer, not a
Kleene-closure construction on either concrete action alphabet. Define
the abstract enabled-response sets

\[
E^{\ast}_{\mathfrak{S}}(m) = \{(i,\alpha(a)) : (i,a) \in E_{\mathfrak{S}}(m)\},
\]

\[
E^{\ast}_{\mathfrak{S}'}(m') = \{(i,\alpha'(a')) : (i,a') \in E_{\mathfrak{S}'}(m')\}.
\]

These abstract enabled-response sets record the labels selected by
enabled agents after abstraction into \(\mathcal{A}^{\ast}\). They are
the comparison observables; later dynamic claims additionally require
the relevant state-action pairs to lie in the admissibility domains of
the update operators.

We say that \(\mathfrak{S}\) and \(\mathfrak{S}'\) are
response-equivalent relative to representation maps

\[
\rho : \mathcal{M} \to \mathcal{C}, \qquad \rho' : \mathcal{M}' \to \mathcal{C}
\]

if

\[
\rho(m) = \rho'(m') \implies E^{\ast}_{\mathfrak{S}}(m) = E^{\ast}_{\mathfrak{S}'}(m')
\]

for all \(m \in \mathcal{M}\) and \(m' \in \mathcal{M}'\).

In words, two media are response-equivalent when the same represented
coordination situation induces the same enabled agent-response structure
after abstraction into a common response alphabet.

Definition 6.2. A family of feature maps
\(\Pi_i : \mathcal{M} \to \mathcal{W}_i\) is response-sufficient for
\(\mathfrak{S}\) if for each agent \(i\) there exist maps

\[
\overline{\beta}_i : \mathcal{W}_i \to \{0,1\}, \qquad \overline{\Gamma}_i^{\ast} : \mathcal{W}_i \to \mathcal{A}^{\ast} \cup \{\bot\}
\]

such that, writing

\[
\beta_i = P_i \circ O_i, \qquad \alpha^{\bot}(a) = \alpha(a) \text{ for } a \in \mathcal{A}, \qquad \alpha^{\bot}(\bot) = \bot,
\]

and

\[
\Gamma_i^{\ast} = \alpha^{\bot} \circ \Gamma_i,
\]

one has

\[
\beta_i(m) = \overline{\beta}_i(\Pi_i(m)), \qquad \Gamma_i^{\ast}(m) = \overline{\Gamma}_i^{\ast}(\Pi_i(m))
\]

for all \(m \in \mathcal{M}\). In that case, \(\Pi_i(m)\) records medium
features sufficient to determine whether agent \(i\) is enabled and, if
enabled, which abstract response it selects. We do not require \(\Pi_i\)
to be minimal: only response-sufficient.

Definition 6.3. For systems \(\mathfrak{S}\) and \(\mathfrak{S}'\) with
abstraction maps \(\alpha\) and \(\alpha'\), write

\[
\beta'_i = P'_i \circ O'_i, \qquad (\alpha')^{\bot}(a') = \alpha'(a') \text{ for } a' \in \mathcal{A}', \qquad (\alpha')^{\bot}(\bot) = \bot,
\]

and

\[
(\Gamma'_i)^{\ast} = (\alpha')^{\bot} \circ \Gamma'_i.
\]

Then the agentwise abstract response signatures are

\[
\Sigma_i(m) = \bigl(\beta_i(m), \Gamma_i^{\ast}(m)\bigr), \qquad \Sigma'_i(m') = \bigl(\beta'_i(m'), (\Gamma'_i)^{\ast}(m')\bigr).
\]

By the standing gating convention (Section 3.2), whenever
\(\beta_i(m) = 0\) one has
\(\Gamma_i^{\ast}(m) = \alpha^{\bot}(\bot) = \bot\), so the signature
reduces to \((0,\bot)\). In particular, signature comparison in disabled
states carries no latent response-label information.

A pair of representation maps \((\rho,\rho')\) is observation-faithful
if

\[
\rho(m) = \rho'(m') \implies \Sigma_i(m) = \Sigma'_i(m') \quad \text{for all } i \in \mathcal{I}.
\]

Definition 6.4. A common coordination space \(\mathcal{C}\) together
with representation maps

\[
\rho : \mathcal{M} \to \mathcal{C}, \qquad \rho' : \mathcal{M}' \to \mathcal{C}
\]

is response-adequate for \(\mathfrak{S}\) and \(\mathfrak{S}'\) if there
exist decoders

\[
\Lambda_i : \mathcal{C} \to \{0,1\} \times (\mathcal{A}^{\ast} \cup \{\bot\})
\]

such that

\[
\Sigma_i = \Lambda_i \circ \rho, \qquad \Sigma'_i = \Lambda_i \circ \rho'
\]

for every agent \(i\). A response-adequate coordination space therefore
carries enough information to reconstruct the abstract enabled-response
signature of each agent in both media.

Definition 6.5. A pair of representation maps \((\rho,\rho')\) is
update-faithful relative to action abstraction maps
\(\alpha : \mathcal{A} \to \mathcal{A}^{\ast}\) and
\(\alpha' : \mathcal{A}' \to \mathcal{A}^{\ast}\) if, whenever states
\(m \in \mathcal{M}\) and \(m' \in \mathcal{M}'\) satisfy
\(\rho(m) = \rho'(m')\), and for some agent \(i\) one has

\[
\Gamma_i(m) = a, \qquad \Gamma'_i(m') = a', \qquad (m,a) \in \operatorname{Adm}, \qquad (m',a') \in \operatorname{Adm}',
\]

with

\[
\alpha(a) = \alpha'(a'),
\]

one has

\[
\rho(U(m,a)) = \rho'(U'(m',a')).
\]

An update-faithful representation therefore intertwines one-step medium
evolution after abstraction: matched abstract actions from equal
represented states remain equal after the update. The definition is
conditional on dynamic admissibility by design: response signatures may
be compared statewise even when some selected responses are not
executable updates, but update-faithfulness speaks only about selected
responses that also lie in the update domains.

\hypertarget{limits-of-the-present-equivalence-notion}{%
\subsection{6.3 Limits of the present equivalence
notion}\label{limits-of-the-present-equivalence-notion}}

Definitions 6.1--6.5 are deliberately weak. They do not require full
pathwise bisimulation, preservation of cost, latency, or failure
semantics, or a complete concurrency theory. They are comparison tools,
not a finished equivalence theory. Their purpose here is to identify the
response-relevant object that later work must strengthen. Section 7 adds
a one-step obstruction theorem for fiber-splitting actions, repair lower
bounds forced by response-fiber classes, and a finite serial closure of
the one-step matched-update result under an explicit matched-action
relation.

\hypertarget{structural-results-metadata-refinement-theorems-a-one-step-obstruction-repair-lower-bounds-and-a-finite-trace-lift}{%
\section{7. Structural Results, Metadata-Refinement Theorems, a One-Step
Obstruction, Repair Lower Bounds, and a Finite Trace
Lift}\label{structural-results-metadata-refinement-theorems-a-one-step-obstruction-repair-lower-bounds-and-a-finite-trace-lift}}

We begin with the structural facts supported directly by the abstract
system class, then develop the adequacy-and-repair spine for metadata
refinements: a sharp one-step obstruction for fiber-splitting actions,
the lower bounds forced by response-fiber classes, and the finite serial
closure of the matched one-step theorem under an explicit execution
model. The main cycle asks when extra medium metadata can be quotiented
away, what any response-adequate comparison must still remember when it
cannot, which refined actions already fail at the one-step action level,
how small any adequate repair can be, and how far the resulting quotient
can be tracked along serial executions.

\begin{namedplain}{Proposition 7.1. Guarded exclusivity}

Assume an exclusive-resource system in which every claim action for
resource \(q\) is admissible only when \(s_q(m) = \mathrm{OPEN}\) and,
when executed, atomically sets \(s_q\) to \(\mathrm{CLAIMED}\) together
with a unique claimant identity. Assume further that release and expiry
actions never assign more than one claimant to the same resource. Under
serial execution of admissible actions, the single-occupancy condition
is invariant.

Proof. Proceed by induction on serial execution length. The base case
holds by assumption on the initial state. For the induction step, let
\(m' = U(m,a)\) for an admissible serial action \(a\), and fix a
resource \(q\). If \(a\) does not affect \(q\), then the owner and
status maps for \(q\) are unchanged, so the single-occupancy condition
carries over from \(m\) to \(m'\). If \(a\) is a claim action on \(q\),
admissibility requires \(s_q(m) = \mathrm{OPEN}\), so no agent occupies
\(q\) in claimed status at state \(m\). Atomicity of the claim update
installs at most one claimant in \(m'\). By assumption, release and
expiry actions also preserve the bound of at most one claimant. Hence
the invariant holds after every serial admissible step.

Remark. This proposition is the abstract mutual-exclusion statement
behind claim, lock, and reservation patterns across several media.

\end{namedplain}

\begin{namedplain}{Proposition 7.2. Closure under product composition}

If each subsystem \(\mathfrak{S}_k\) is admissible, the joint
observation and response operators remain total on the product medium,
and the joint update rule preserves the product state space, then the
product system is again an admissible shared-medium stigmergic system of
the same form.

Proof. The admissibility domain of the product system is specified by
hypothesis. Closure of the joint update rule on the product state space
is also assumed. Totality of each joint observation operator follows
from totality of the coordinate observations because each coordinate
state lies in its own admissible state space. The activation predicates
and response maps in the product system still depend only on the joint
observation made available to the corresponding agent. Hence the
defining admissibility conditions of Section 4.1 hold for the product
system.

Remark. This proposition is the structural reason cross-contract,
cross-dataspace, and cross-medium coordination can be treated
compositionally rather than as unrelated extensions.

\end{namedplain}

\begin{namedremark}{Remark 7.3. Persistence as the special case of decay}

Persistent-trace systems are exactly decaying-trace systems with decay
operator \(D = \mathrm{id}\). If \(D = \mathrm{id}\), then inter-step
evolution is \(m_{t+1} = U(m_t,a_t)\), so traces persist until altered
by later updates. Conversely, any persistent-trace system may be written
in decaying form by taking \(D\) to be the identity on \(\mathcal{M}\).
This places pheromone evaporation, timed expiry, and persistent digital
media inside one operator template.

\end{namedremark}

\hypertarget{abstract-response-comparison}{%
\subsection{7.2 Abstract response
comparison}\label{abstract-response-comparison}}

Notation recalled from Section 6. For any admissible system,
\(\Sigma(m) = (\Sigma_i(m))_{i \in \mathcal{I}}\) denotes the agentwise
abstract response-signature tuple of Definition 6.3, and Lemma 7.4 turns
it into the abstract enabled-response set
\(E_{\mathfrak{S}}^{\ast}(m)\). In the one-step comparison results
below, \(B \subseteq \mathcal{A} \times \widetilde{\mathcal{A}}\) will
denote a model-specific relation of matched actions, and
``update-faithful'' refers to the commutative square of Definition 6.5
on agent-selected matched pairs.

Standing assumptions for Lemma 7.4--Theorem 7.10A. Unless otherwise
stated, the systems being compared share the same agent set
\(\mathcal{I}\), the action abstraction maps \(\alpha\) and \(\alpha'\)
are fixed once and for all, update claims are made only on admissible
state-action pairs in the domains of \(U\) and \(U'\), and equality in
the coordination space \(\mathcal{C}\) means literal equality of
represented states rather than equality modulo an external quotient. The
results of Sections 7.2--7.3 are therefore statewise or one-step
statements. Section 7.4 adds a finite serial matched-trace lift, but not
a full branching or concurrency theory.

\begin{namedplain}{Lemma 7.4. Signature tuples determine abstract enabled-response sets}

Let \(\mathfrak{S}\) be an admissible stigmergic system with action
abstraction map \(\alpha : \mathcal{A} \to \mathcal{A}^{\ast}\). If
\(\Sigma(m) = (\Sigma_i(m))_{i \in \mathcal{I}}\) is the abstract
response-signature tuple of Definition 6.3, then
\(E^{\ast}_{\mathfrak{S}}(m)\) is determined entirely by \(\Sigma(m)\).

Proof. By definition, \(\Sigma_i(m)\) records whether agent \(i\) is
enabled and, when enabled, which abstract response \(\alpha(a)\) it
selects. Therefore

\[
E^{\ast}_{\mathfrak{S}}(m) = \{(i,a^{\ast}) \in \mathcal{I} \times \mathcal{A}^{\ast} : \Sigma_i(m) = (1,a^{\ast})\},
\]

so the full abstract enabled-response set is determined by the signature
tuple.

Remark. The lemma identifies the comparison object used below: not raw
medium state, but the agentwise abstract response signature extracted
from it.

\end{namedplain}

\begin{namedplain}{Theorem 7.5. Response equivalence under observation-faithful representation}

Assumptions. Let \(\mathfrak{S}\) and \(\mathfrak{S}'\) be admissible
stigmergic systems over the common agent set \(\mathcal{I}\) with fixed
action abstraction maps \(\alpha : \mathcal{A} \to \mathcal{A}^{\ast}\)
and \(\alpha' : \mathcal{A}' \to \mathcal{A}^{\ast}\). Assume that
\(\rho : \mathcal{M} \to \mathcal{C}\) and
\(\rho' : \mathcal{M}' \to \mathcal{C}\) are observation-faithful in the
sense of Definition 6.3.

Statement. For all \(m \in \mathcal{M}\) and \(m' \in \mathcal{M}'\),

\[
\rho(m) = \rho'(m') \implies E^{\ast}_{\mathfrak{S}}(m) = E^{\ast}_{\mathfrak{S}'}(m').
\]

Equivalently, \(\mathfrak{S}\) and \(\mathfrak{S}'\) are
response-equivalent relative to \((\rho,\rho',\alpha,\alpha')\).

Proof. If \(\rho(m) = \rho'(m')\), observation-faithfulness gives

\[
\Sigma_i(m) = \Sigma'_i(m') \quad \text{for all } i \in \mathcal{I}.
\]

Lemma 7.4 then implies equality of the two abstract enabled-response
sets.

Scope. The result is statewise: it identifies when represented states
support the same abstract enabled responses, without comparing longer
executions or pathwise simulations.

\end{namedplain}

\begin{namedplain}{Corollary 7.6. Response-adequate coordination spaces imply response equivalence}

Assumptions. Let \(\mathcal{C}\) be response-adequate for admissible
systems \(\mathfrak{S}\) and \(\mathfrak{S}'\) with respect to
representation maps \(\rho\) and \(\rho'\) and fixed abstraction maps
\(\alpha\) and \(\alpha'\).

Statement. Then \(\mathfrak{S}\) and \(\mathfrak{S}'\) are
response-equivalent relative to \((\rho,\rho',\alpha,\alpha')\).

Proof. Response adequacy provides shared decoders \(\Lambda_i\) with

\[
\Sigma_i = \Lambda_i \circ \rho, \qquad \Sigma'_i = \Lambda_i \circ \rho'
\]

for every agent \(i\). Hence equality in \(\mathcal{C}\) implies
equality of all agentwise signatures, so \((\rho,\rho')\) is
observation-faithful. The conclusion follows from Theorem 7.5.

Scope. The corollary pins down what a useful common coordination space
must encode, but adds no dynamic preservation beyond the statewise
comparison of Theorem 7.5.

\end{namedplain}

\begin{namedplain}{Lemma 7.7. The canonical signature maps are response-sufficient}

For every admissible stigmergic system \(\mathfrak{S}\), the family of
maps

\[
\Pi_i = \Sigma_i : \mathcal{M} \to \{0,1\} \times (\mathcal{A}^{\ast} \cup \{\bot\})
\]

is response-sufficient in the sense of Definition 6.2.

Proof. Take \(\overline{\beta}_i\) and \(\overline{\Gamma}_i^{\ast}\) to
be the coordinate projections on
\(\{0,1\} \times (\mathcal{A}^{\ast} \cup \{\bot\})\). Then Definition
6.2 holds identically.

Scope. The lemma provides a canonical choice of response-sufficient
features intrinsic to the abstract comparison problem.

\end{namedplain}

\hypertarget{metadata-refinements-aligned-fibers-and-canonical-repairs}{%
\subsection{7.3 Metadata refinements, aligned fibers, and canonical
repairs}\label{metadata-refinements-aligned-fibers-and-canonical-repairs}}

The next results isolate a recurrent comparison situation. A refined
medium may carry extra metadata over a coarser base medium, such as
timestamps, propagation markers, or version data. The question is when
the forgetting map to the coarser medium preserves the abstract response
signatures and, when it does not, what the coarsest adequate repair must
remember.

Given a surjective projection
\(\pi : \widetilde{\mathcal{M}} \to \mathcal{M}\), write

\[
\Sigma(m) = (\Sigma_i(m))_{i \in \mathcal{I}}, \qquad \widetilde{\Sigma}(\widetilde{m}) = (\widetilde{\Sigma}_i(\widetilde{m}))_{i \in \mathcal{I}},
\]

and for each coarse state \(m \in \mathcal{M}\) define the refined
response-signature set above \(m\) by

\[
F_{\pi}(m) = \{\widetilde{\Sigma}(\widetilde{m}) : \widetilde{m} \in \widetilde{\mathcal{M}},\; \pi(\widetilde{m}) = m\}.
\]

We call the fiber \(\pi^{-1}(m)\) response-aligned if

\[
F_{\pi}(m) = \{\Sigma(m)\},
\]

or equivalently if \(\widetilde{\Sigma}(\widetilde{m}) = \Sigma(m)\) for
every \(\widetilde{m} \in \pi^{-1}(m)\).

We also define the response-fiber equivalence relation on refined states
by

\[
\widetilde{m}_1 \sim_{\mathrm{resp}} \widetilde{m}_2
\quad \Longleftrightarrow \quad
\pi(\widetilde{m}_1) = \pi(\widetilde{m}_2)
\text{ and }
\widetilde{\Sigma}(\widetilde{m}_1) = \widetilde{\Sigma}(\widetilde{m}_2).
\]

Its equivalence classes are the response-fiber classes that the
canonical repair below must remember.

\begin{namedplain}{Theorem 7.8. Quotient adequacy criterion for metadata refinements}

Assumptions. Let

\[
\mathfrak{S} = \langle \mathcal{M}, \mathcal{A}, \mathcal{I}, U, \{O_i\}, \{P_i\}, \{R_i\} \rangle,
\qquad
\widetilde{\mathfrak{S}} = \langle \widetilde{\mathcal{M}}, \widetilde{\mathcal{A}}, \mathcal{I}, \widetilde{U}, \{\widetilde{O}_i\}, \{\widetilde{P}_i\}, \{\widetilde{R}_i\} \rangle
\]

be admissible stigmergic systems with fixed action abstraction maps

\[
\alpha : \mathcal{A} \to \mathcal{A}^{\ast}, \qquad \widetilde{\alpha} : \widetilde{\mathcal{A}} \to \mathcal{A}^{\ast},
\]

and let \(\pi : \widetilde{\mathcal{M}} \to \mathcal{M}\) be surjective.

Statement. The following are equivalent:

\begin{enumerate}
\def\labelenumi{\arabic{enumi}.}
\tightlist
\item
  The coarse medium \(\mathcal{M}\) is response-adequate for
  \(\mathfrak{S}\) and \(\widetilde{\mathfrak{S}}\) via the
  representation maps \[
  \rho = \mathrm{id}_{\mathcal{M}}, \qquad \widetilde{\rho} = \pi.
  \]
\item
  Every fiber of \(\pi\) is response-aligned.
\end{enumerate}

In either case, \(\mathfrak{S}\) and \(\widetilde{\mathfrak{S}}\) are
response-equivalent relative to
\((\rho,\widetilde{\rho},\alpha,\widetilde{\alpha})\).

Proof. Assume (1). Response adequacy provides decoders
\(\Lambda_i : \mathcal{M} \to \{0,1\} \times (\mathcal{A}^{\ast} \cup \{\bot\})\)
such that

\[
\Sigma_i = \Lambda_i \circ \rho = \Lambda_i, \qquad \widetilde{\Sigma}_i = \Lambda_i \circ \widetilde{\rho} = \Lambda_i \circ \pi.
\]

Hence for every refined state \(\widetilde{m}\) and every agent \(i\),

\[
\widetilde{\Sigma}_i(\widetilde{m}) = \Lambda_i(\pi(\widetilde{m})) = \Sigma_i(\pi(\widetilde{m})).
\]

So every fiber is response-aligned. Conversely, assume (2). Define
\(\Lambda_i : \mathcal{M} \to \{0,1\} \times (\mathcal{A}^{\ast} \cup \{\bot\})\)
by \(\Lambda_i(m) = \Sigma_i(m)\). Then

\[
\Sigma_i = \Lambda_i \circ \rho
\]

holds trivially, and response alignment gives

\[
\widetilde{\Sigma}_i(\widetilde{m}) = \Sigma_i(\pi(\widetilde{m})) = \Lambda_i(\widetilde{\rho}(\widetilde{m}))
\]

for every refined state \(\widetilde{m}\). Thus \(\mathcal{M}\) is
response-adequate for the pair via \((\rho,\widetilde{\rho})\). The
final claim follows from Corollary 7.6.

Scope. This is the boundary result for response-invisible metadata: the
coarse medium suffices exactly when each quotient fiber carries the same
abstract response signature as its coarse image.

\end{namedplain}

\begin{namedplain}{Theorem 7.9. One-step update preservation for metadata-compatible refinements}

Definition. Under the hypotheses of Theorem 7.8, a relation
\(B \subseteq \mathcal{A} \times \widetilde{\mathcal{A}}\) is called
quotient-compatible if for every \((a,\widetilde{a}) \in B\) and every
refined state-action pair \((\widetilde{m},\widetilde{a})\) in the
domain of \(\widetilde{U}\),

\[
(\pi(\widetilde{m}),a) \text{ lies in the domain of } U,
\qquad
\widetilde{\alpha}(\widetilde{a}) = \alpha(a)
\]

and

\[
\pi(\widetilde{U}(\widetilde{m},\widetilde{a})) = U(\pi(\widetilde{m}),a).
\]

Assumptions. Assume the hypotheses of Theorem 7.8 and let
\(B \subseteq \mathcal{A} \times \widetilde{\mathcal{A}}\) be a
quotient-compatible relation of matched actions.

Statement. The representation pair \((\mathrm{id}_{\mathcal{M}},\pi)\)
is update-faithful in the sense of Definition 6.5 on any agent-selected
matched action pair drawn from \(B\). In particular, if
\((a,\widetilde{a}) \in B\), \(\pi(\widetilde{m}) = m\), and the refined
pair \((\widetilde{m},\widetilde{a})\) lies in the domain of
\(\widetilde{U}\), then \((m,a)\) lies in the domain of \(U\) and

\[
\pi(\widetilde{U}(\widetilde{m},\widetilde{a})) = U(m,a).
\]

If, in addition, every fiber of \(\pi\) is response-aligned, then the
successor states are response-equivalent by Theorem 7.8 and Corollary
7.6.

Proof. Fix \((a,\widetilde{a}) \in B\) and a refined state
\(\widetilde{m}\) such that \((\widetilde{m},\widetilde{a})\) lies in
the domain of \(\widetilde{U}\), and write \(m = \pi(\widetilde{m})\).
Because \(B\) is quotient-compatible, one has

\[
\widetilde{\alpha}(\widetilde{a}) = \alpha(a)
\]

and

\[
\pi(\widetilde{U}(\widetilde{m},\widetilde{a})) = U(\pi(\widetilde{m}),a) = U(m,a).
\]

Since \(\rho = \mathrm{id}_{\mathcal{M}}\) and
\(\widetilde{\rho} = \pi\), this is exactly the update-faithfulness
condition of Definition 6.5 on any agent-selected matched pair from
\(B\). If every fiber of \(\pi\) is response-aligned, then Theorem 7.8
makes \(\mathcal{M}\) itself response-adequate for the pair, so
Corollary 7.6 yields response equivalence of the successor states.

Scope. Here the quotient-compatible matching condition is packaged into
its dynamic consequence: one-step matched evolution yields a projected
coarse successor, and under fiber alignment the successor states are
response-equivalent.

Remark. Constructing the matched-action relation \(B\) is a separate
modeling obligation. In practice, \(B\) is obtained by fixing the
visible action skeleton that survives the quotient and then retaining
exactly those refined actions whose projected successors commute with
\(\pi\). Section 8.4 follows this recipe for unconditional writes by
matching visible key, payload, and phase, while Section 8.4A shows that
guarded freshness actions fail the same commutativity test and therefore
fall outside any such \(B\). Theorems 7.9 and 7.11 prove the
consequences once such a family of matched actions is given, but the
manuscript still does not provide a canonical construction principle,
maximality theorem, or decision procedure for obtaining \(B\) in
general.

\end{namedplain}

\begin{namedplain}{Theorem 7.9A. Fiber-splitting refined actions obstruct uniform coarse matching}

Assumptions. Assume the standing hypotheses of Section 7.2. Let
\(\widetilde{a} \in \widetilde{\mathcal{A}}\) and suppose there exist
refined states
\(\widetilde{m}_1,\widetilde{m}_2 \in \widetilde{\mathcal{M}}\) such
that both pairs

\[
(\widetilde{m}_1,\widetilde{a}), \qquad (\widetilde{m}_2,\widetilde{a})
\]

lie in the domain of \(\widetilde{U}\),

\[
\pi(\widetilde{m}_1) = \pi(\widetilde{m}_2) = m,
\]

and

\[
\pi(\widetilde{U}(\widetilde{m}_1,\widetilde{a})) \neq \pi(\widetilde{U}(\widetilde{m}_2,\widetilde{a})).
\]

Statement. There is no coarse action \(a \in \mathcal{A}\) such that the
quotient-compatible matching condition of Theorem 7.9 holds for the pair
\((a,\widetilde{a})\) at both refined states. Consequently no
quotient-compatible matched-action relation
\(B \subseteq \mathcal{A} \times \widetilde{\mathcal{A}}\) can contain a
pair matching \(\widetilde{a}\) uniformly on the whole fiber
\(\pi^{-1}(m)\).

Proof. If such a coarse action \(a\) existed, then the
quotient-compatible condition of Theorem 7.9 would give

\[
\pi(\widetilde{U}(\widetilde{m}_1,\widetilde{a})) = U(\pi(\widetilde{m}_1),a) = U(m,a) = U(\pi(\widetilde{m}_2),a) = \pi(\widetilde{U}(\widetilde{m}_2,\widetilde{a})),
\]

contrary to the hypothesis that the two projected successors are
different.

Scope. As the obstruction counterpart of Theorem 7.9, this gives a
one-step non-liftability criterion by contraposing the commutative
square. Whenever a refined action splits projected successors inside a
single quotient fiber, hidden metadata blocks any uniform coarse
representation of that action.

Remark. The converse fails in general. Even if
\(\pi \circ \widetilde{U}(\cdot,\widetilde{a})\) is constant on a
quotient fiber, a uniform coarse match may still fail because
\(\mathcal{A}\) may contain no coarse action with that projected
successor, or because any such coarse action may carry the wrong
abstract label under \(\alpha\). Theorem 7.9A therefore records a
sufficient obstruction, not a full characterization of matchability.

The repair theorem below is stated on full comparison spaces, not only
on represented images. Clause (3) rules out spurious comparison points
by requiring every decoded signature to be realizable over its projected
coarse state, and the morphism notion records exactly the structure that
the universal property must preserve: coarse projection together with
decoded abstract response signatures.

Definition 7.9B. A response-adequate repair of the quotient comparison
determined by \(\pi\) is a tuple

\[
(\mathcal{C},\rho,\widetilde{\rho},p,\{\Lambda_i\}_{i \in \mathcal{I}})
\]

such that

\begin{enumerate}
\def\labelenumi{\arabic{enumi}.}
\tightlist
\item
  \(\mathcal{C}\) is response-adequate for \(\mathfrak{S}\) and
  \(\widetilde{\mathfrak{S}}\) via \((\rho,\widetilde{\rho})\),
  equivalently \[
  \Sigma_i = \Lambda_i \circ \rho, \qquad \widetilde{\Sigma}_i = \Lambda_i \circ \widetilde{\rho}
  \] for every agent \(i\).
\item
  The projection \(p : \mathcal{C} \to \mathcal{M}\) satisfies \[
  p \circ \rho = \mathrm{id}_{\mathcal{M}}, \qquad p \circ \widetilde{\rho} = \pi.
  \]
\item
  Writing \(\Lambda(c) = (\Lambda_i(c))_{i \in \mathcal{I}}\), one has
  \[
  \Lambda(c) \in F_{\pi}(p(c)) \cup \{\Sigma(p(c))\}
  \] for every \(c \in \mathcal{C}\).
\end{enumerate}

Write

\[
\mathcal{R} = (\mathcal{C},\rho,\widetilde{\rho},p,\{\Lambda_i\}), \qquad \mathcal{R}' = (\mathcal{C}',\rho',\widetilde{\rho}',p',\{\Lambda_i'\}).
\]

A morphism of response-adequate repairs from \(\mathcal{R}\) to
\(\mathcal{R}'\) is a map \(H : \mathcal{C} \to \mathcal{C}'\) such that

\[
p' \circ H = p, \qquad \Lambda_i' \circ H = \Lambda_i \quad \text{for all } i \in \mathcal{I}.
\]

Remark on Definition 7.9B(3). Clause (3) is a realizability restriction,
not decoration: it forbids comparison points \(c \in \mathcal{C}\) whose
decoded signature \(\Lambda(c)\) is not realized over the projected
coarse state \(p(c)\). Every such \(\Lambda(c)\) must already arise
either as the coarse signature \(\Sigma(p(c))\) or as the refined
signature of some preimage in \(\pi^{-1}(p(c))\), i.e.~as an element of
\(F_{\pi}(p(c)) \cup \{\Sigma(p(c))\}\). Without this clause a repair
could pair coarse states with arbitrary signature ghosts, defeating the
universal property of Theorem 7.10. The theorem is therefore universal
for this deliberately realized notion of repair, not for every
imaginable covering of the quotient.

\end{namedplain}

\begin{namedplain}{Theorem 7.10. Canonical coarsest repair by response-fiber classes}

Assumptions. Let \(\mathfrak{S}\), \(\widetilde{\mathfrak{S}}\),
\(\alpha\), \(\widetilde{\alpha}\), and \(\pi\) be as above. Define

\[
\mathcal{C}_{\min} = \{(m,\sigma) : m \in \mathcal{M},\; \sigma \in F_{\pi}(m) \cup \{\Sigma(m)\}\},
\]

with representation maps, coarse projection, and canonical decoders

\[
\rho_{\min}(m) = (m,\Sigma(m)), \qquad \widetilde{\rho}_{\min}(\widetilde{m}) = (\pi(\widetilde{m}),\widetilde{\Sigma}(\widetilde{m})),
\]

\[
p_{\min}(m,\sigma) = m, \qquad \Lambda_i^{\min}(m,\sigma) = \sigma_i.
\]

Statement.

\begin{enumerate}
\def\labelenumi{\arabic{enumi}.}
\tightlist
\item
  The tuple
  \((\mathcal{C}_{\min},\rho_{\min},\widetilde{\rho}_{\min},p_{\min},\{\Lambda_i^{\min}\}_{i \in \mathcal{I}})\)
  is a response-adequate repair of the quotient comparison determined by
  \(\pi\).
\item
  For refined states
  \(\widetilde{m}_1,\widetilde{m}_2 \in \widetilde{\mathcal{M}}\), \[
  \widetilde{\rho}_{\min}(\widetilde{m}_1) = \widetilde{\rho}_{\min}(\widetilde{m}_2)
  \quad \Longleftrightarrow \quad
  \widetilde{m}_1 \sim_{\mathrm{resp}} \widetilde{m}_2.
  \]
\item
  For every response-adequate repair
  \((\mathcal{C},\rho,\widetilde{\rho},p,\{\Lambda_i\}_{i \in \mathcal{I}})\)
  of the quotient comparison determined by \(\pi\), there exists a
  unique morphism of response-adequate repairs \[
  H : \mathcal{C} \to \mathcal{C}_{\min}, \qquad H(c) = \bigl(p(c), (\Lambda_i(c))_{i \in \mathcal{I}}\bigr).
  \] In particular, \[
  H \circ \rho = \rho_{\min}, \qquad H \circ \widetilde{\rho} = \widetilde{\rho}_{\min}.
  \]
\end{enumerate}

Proof. For (1), the identities

\[
\Lambda_i^{\min}(\rho_{\min}(m)) = \Sigma_i(m), \qquad \Lambda_i^{\min}(\widetilde{\rho}_{\min}(\widetilde{m})) = \widetilde{\Sigma}_i(\widetilde{m})
\]

hold for every agent \(i\), so \(\mathcal{C}_{\min}\) is
response-adequate by Definition 6.4. Also

\[
p_{\min}(\rho_{\min}(m)) = m, \qquad p_{\min}(\widetilde{\rho}_{\min}(\widetilde{m})) = \pi(\widetilde{m}),
\]

and if \(c = (m,\sigma) \in \mathcal{C}_{\min}\), then by definition
\(\sigma \in F_{\pi}(m) \cup \{\Sigma(m)\} = F_{\pi}(p_{\min}(c)) \cup \{\Sigma(p_{\min}(c))\}\).
Thus all three clauses of the repair definition hold.

For (2), one has

\[
\widetilde{\rho}_{\min}(\widetilde{m}_1) = \widetilde{\rho}_{\min}(\widetilde{m}_2)
\]

if and only if the ordered pairs

\[
(\pi(\widetilde{m}_1),\widetilde{\Sigma}(\widetilde{m}_1))
\quad \text{and} \quad
(\pi(\widetilde{m}_2),\widetilde{\Sigma}(\widetilde{m}_2))
\]

are equal, which is equivalent to the two conditions in the definition
of \(\sim_{\mathrm{resp}}\).

For (3), let \((\mathcal{C},\rho,\widetilde{\rho},p,\{\Lambda_i\})\) be
any response-adequate repair. For each \(c \in \mathcal{C}\), clause (3)
of Definition 7.9B gives

\[
(\Lambda_i(c))_{i \in \mathcal{I}} \in F_{\pi}(p(c)) \cup \{\Sigma(p(c))\},
\]

so \(H(c) \in \mathcal{C}_{\min}\) and the displayed formula defines a
map \(H : \mathcal{C} \to \mathcal{C}_{\min}\). By construction,

\[
p_{\min}(H(c)) = p(c), \qquad \Lambda_i^{\min}(H(c)) = \Lambda_i(c)
\]

for every \(c \in \mathcal{C}\) and every agent \(i\), so \(H\) is a
morphism of repairs. If \(G : \mathcal{C} \to \mathcal{C}_{\min}\) is
any other morphism of repairs, then for every \(c \in \mathcal{C}\),

\[
G(c) = \bigl(p_{\min}(G(c)), (\Lambda_i^{\min}(G(c)))_{i \in \mathcal{I}}\bigr) = \bigl(p(c), (\Lambda_i(c))_{i \in \mathcal{I}}\bigr) = H(c),
\]

because points of \(\mathcal{C}_{\min}\) are exactly coarse states
paired with response signatures. Hence \(H\) is unique. Finally,

\[
H(\rho(m)) = \bigl(p(\rho(m)), (\Lambda_i(\rho(m)))_{i \in \mathcal{I}}\bigr) = (m,\Sigma(m)) = \rho_{\min}(m),
\]

and similarly
\(H(\widetilde{\rho}(\widetilde{m})) = \widetilde{\rho}_{\min}(\widetilde{m})\).

Scope. The theorem upgrades the earlier separation result to a genuine
global universal property. The repaired space \(\mathcal{C}_{\min}\)
remembers exactly the response-fiber class and is the unique coarsest
response-adequate repair preserving coarse projection and decoded
signatures. Equivalently, \(\mathcal{C}_{\min}\) is a terminal object in
the category of response-adequate repairs of the quotient comparison
determined by \(\pi\), where morphisms are as introduced in the
paragraph preceding Theorem 7.10. This is the \emph{qualitative} sense
in which the repair is canonical. The \emph{quantitative} sense,
recorded separately in Theorem 7.10A, is a fiberwise lower bound on the
size of any response-adequate repair; the two notions of ``coarsest''
are independent and complementary.

\end{namedplain}

\begin{namedplain}{Theorem 7.10A. Fiberwise lower bounds forced by response-fiber classes}

Assumptions. Let

\[
(\mathcal{C},\rho,\widetilde{\rho},p,\{\Lambda_i\}_{i \in \mathcal{I}})
\]

be any response-adequate repair of the quotient comparison determined by
\(\pi\).

Statement.

\begin{enumerate}
\def\labelenumi{\arabic{enumi}.}
\tightlist
\item
  For every coarse state \(m \in \mathcal{M}\), \[
  |p^{-1}(m)| \ge |F_{\pi}(m) \cup \{\Sigma(m)\}|.
  \]
\item
  If \(\mathcal{M}\) is finite, then \[
  |\mathcal{C}| \ge \sum_{m \in \mathcal{M}} |F_{\pi}(m) \cup \{\Sigma(m)\}|.
  \]
\item
  The canonical repair of Theorem 7.10 attains equality fiberwise: \[
  |p_{\min}^{-1}(m)| = |F_{\pi}(m) \cup \{\Sigma(m)\}|
  \] for every \(m \in \mathcal{M}\). In particular, if \(\mathcal{M}\)
  is finite then \[
  |\mathcal{C}_{\min}| = \sum_{m \in \mathcal{M}} |F_{\pi}(m) \cup \{\Sigma(m)\}|.
  \]
\end{enumerate}

Proof. Fix \(m \in \mathcal{M}\). For each signature tuple

\[
\tau \in F_{\pi}(m) \cup \{\Sigma(m)\},
\]

choose a point \(c_{\tau} \in p^{-1}(m)\) as follows. If
\(\tau = \Sigma(m)\), set

\[
c_{\tau} = \rho(m).
\]

Otherwise choose a refined witness
\(\widetilde{m}_{\tau} \in \pi^{-1}(m)\) with

\[
\widetilde{\Sigma}(\widetilde{m}_{\tau}) = \tau
\]

and set

\[
c_{\tau} = \widetilde{\rho}(\widetilde{m}_{\tau}).
\]

In both cases the repair identities give

\[
p(c_{\tau}) = m.
\]

If \(c_{\tau_1} = c_{\tau_2}\) for two signatures
\(\tau_1 = ((\tau_1)_i)_{i \in \mathcal{I}}\) and
\(\tau_2 = ((\tau_2)_i)_{i \in \mathcal{I}}\), then for every agent
\(i\),

\[
(\tau_1)_i = \Lambda_i(c_{\tau_1}) = \Lambda_i(c_{\tau_2}) = (\tau_2)_i,
\]

so \(\tau_1 = \tau_2\). Thus

\[
\tau \longmapsto c_{\tau}
\]

is an injection from \(F_{\pi}(m) \cup \{\Sigma(m)\}\) into
\(p^{-1}(m)\), proving (1). Since the fibers of \(p\) over distinct
coarse states are disjoint and partition \(\mathcal{C}\), summing (1)
over \(m \in \mathcal{M}\) gives (2) whenever \(\mathcal{M}\) is finite.

For (3), one has by definition

\[
p_{\min}^{-1}(m) = \{(m,\sigma) : \sigma \in F_{\pi}(m) \cup \{\Sigma(m)\}\},
\]

so the displayed fiberwise equality is immediate, and the global
equality follows by summing over \(m\) when \(\mathcal{M}\) is finite.

Scope. This is the obstruction form of Theorem 7.10: the canonical
repair is both universal and minimal, since no response-adequate repair
can compress a coarse state below the number of realized response-fiber
classes that must be distinguished there.

\end{namedplain}

\hypertarget{finite-matched-traces-under-serial-execution}{%
\subsection{7.4 Finite matched traces under serial
execution}\label{finite-matched-traces-under-serial-execution}}

The next result should be read as the finite serial closure of Theorem
7.9 under the minimal execution model below. Its role is to make the
trace-level claim precise, not to introduce a second theorem of the same
depth as Theorems 7.8 or 7.10. To do that without importing a full
concurrent semantics, we now fix the minimal serial execution model
already implicit in Proposition 7.1. A finite execution fragment of an
admissible system \(\mathfrak{S}\) is a sequence

\[
m_0 \xrightarrow{a_0} m_1 \xrightarrow{a_1} \cdots \xrightarrow{a_{n-1}} m_n
\]

such that each pair \((m_t,a_t)\) lies in the domain of \(U\) and

\[
m_{t+1} = U(m_t,a_t)
\]

for \(t = 0,\dots,n-1\).

Given a matched-action relation
\(B \subseteq \mathcal{A} \times \widetilde{\mathcal{A}}\), a coarse
fragment

\[
m_0 \xrightarrow{a_0} \cdots \xrightarrow{a_{n-1}} m_n
\]

and a refined fragment

\[
\widetilde{m}_0 \xrightarrow{\widetilde{a}_0} \cdots \xrightarrow{\widetilde{a}_{n-1}} \widetilde{m}_n
\]

of the same length are called \(B\)-matched if
\((a_t,\widetilde{a}_t) \in B\) for every step \(t\). The corresponding
abstract response traces are the state-indexed sequences
\((\Sigma(m_t))_{t=0}^{n}\) and
\((\widetilde{\Sigma}(\widetilde{m}_t))_{t=0}^{n}\), or equivalently by
Lemma 7.4 the enabled-response traces
\((E_{\mathfrak{S}}^{\ast}(m_t))_{t=0}^{n}\) and
\((E_{\widetilde{\mathfrak{S}}}^{\ast}(\widetilde{m}_t))_{t=0}^{n}\).

\begin{namedplain}{Theorem 7.11. Finite serial closure of matched one-step preservation}

Assumptions. Assume the hypotheses of Theorem 7.9. Let

\[
\widetilde{m}_0 \xrightarrow{\widetilde{a}_0} \widetilde{m}_1 \xrightarrow{\widetilde{a}_1} \cdots \xrightarrow{\widetilde{a}_{n-1}} \widetilde{m}_n
\]

be a finite execution fragment of \(\widetilde{\mathfrak{S}}\), and
choose coarse actions \(a_0,\dots,a_{n-1} \in \mathcal{A}\) such that
\((a_t,\widetilde{a}_t) \in B\) for every \(t\). Set

\[
m_0 = \pi(\widetilde{m}_0).
\]

The proof below establishes by induction that each pair \((m_t,a_t)\)
lies in the domain of \(U\) via Theorem 7.9, so the recursion
\(m_{t+1} = U(m_t,a_t)\) for \(t = 0,\dots,n-1\) is well-defined.

Statement.

\begin{enumerate}
\def\labelenumi{\arabic{enumi}.}
\tightlist
\item
  The sequence \[
  m_0 \xrightarrow{a_0} m_1 \xrightarrow{a_1} \cdots \xrightarrow{a_{n-1}} m_n
  \] is a finite execution fragment of \(\mathfrak{S}\).
\item
  For every \(t = 0,\dots,n\), \[
  \pi(\widetilde{m}_t) = m_t.
  \]
\item
  If every fiber of \(\pi\) is response-aligned, then for every
  \(t = 0,\dots,n\), \[
  \widetilde{\Sigma}(\widetilde{m}_t) = \Sigma(m_t)
  \] and hence \[
  E_{\widetilde{\mathfrak{S}}}^{\ast}(\widetilde{m}_t) = E_{\mathfrak{S}}^{\ast}(m_t).
  \]
\end{enumerate}

Equivalently, every finite refined serial fragment whose steps are
matched by \(B\) projects to a unique coarse fragment, and under fiber
alignment the two fragments have the same abstract enabled-response
trace.

Proof. We prove (1) and (2) simultaneously by induction on \(t\). The
base case is immediate from the definition
\(m_0 = \pi(\widetilde{m}_0)\). For the induction step, assume
\(\pi(\widetilde{m}_t) = m_t\). Since

\[
\widetilde{m}_t \xrightarrow{\widetilde{a}_t} \widetilde{m}_{t+1}
\]

is a step of a refined execution fragment, the pair
\((\widetilde{m}_t,\widetilde{a}_t)\) lies in the domain of
\(\widetilde{U}\). Because \((a_t,\widetilde{a}_t) \in B\), Theorem 7.9
gives that \((m_t,a_t)\) lies in the domain of \(U\) and

\[
\pi(\widetilde{m}_{t+1})
=
\pi(\widetilde{U}(\widetilde{m}_t,\widetilde{a}_t))
=
U(\pi(\widetilde{m}_t),a_t)
=
U(m_t,a_t)
=
m_{t+1}.
\]

Thus the coarse sequence is an execution fragment and the projection
identity holds at step \(t+1\). Here the coarse admissibility of
\((m_t,a_t)\) comes from the quotient-compatibility of \(B\), not from
the induction hypothesis alone. This proves (1) and (2).

For (3), assume every fiber of \(\pi\) is response-aligned. Then Theorem
7.8 makes \(\mathcal{M}\) response-adequate for the pair via
\((\mathrm{id}_{\mathcal{M}},\pi)\), so

\[
\widetilde{\Sigma}(\widetilde{m}_t) = \Sigma(\pi(\widetilde{m}_t)) = \Sigma(m_t)
\]

for every \(t\). Lemma 7.4 then yields equality of the abstract
enabled-response sets at each step. The uniqueness claim follows because
\(m_0\) is fixed by \(\pi(\widetilde{m}_0)\) and the coarse successor at
each step is determined uniquely by the deterministic update operator
\(U\) on admissible pairs.

Scope. This is an inductive closure of Theorem 7.9, not a new invariant
theorem. It covers only finite serial fragments whose steps are
explicitly matched by \(B\); the intended applications are turn-based or
otherwise single-threaded shared-medium interactions. Trace equality
additionally requires fiber alignment. Branching structure, unmatched
actions, fairness assumptions, and concurrent interleavings are not
addressed.

\end{namedplain}

\hypertarget{worked-dataspace-instantiation}{%
\section{8. Worked Dataspace
Instantiation}\label{worked-dataspace-instantiation}}

We now work out one explicit pair of media rather than surveying many in
parallel. The pair is used to check the abstract hypotheses and to
locate the points where the quotient succeeds, matched writes preserve
it for one step, freshness-conditional actions block a uniform coarse
representative, repair size is forced by response fibers, and
preservation lifts along finite serial executions. Sections 8.3--8.6
realize Theorems 7.8--7.11 together with the dynamic obstruction after
Theorem 7.9 and the repair lower-bound result after Theorem 7.10.

\hypertarget{tuple-space-or-shared-dataspace-medium}{%
\subsection{8.1 Tuple-space or shared-dataspace
medium}\label{tuple-space-or-shared-dataspace-medium}}

Fix a finite key set \(K\), a payload set \(P\), and a phase set
\(\Phi\). A tuple-space or shared-dataspace state is a pair

\[
x = (s,\varphi)
\]

where \(s : K \rightharpoonup P\) is a finite partial payload store and
\(\varphi : K \to \Phi\) records any protocol phase or status tag
associated with each key. Let \(\mathcal{M}_{\mathrm{ts}}\) be the class
of such states.

To keep the worked instantiation theorem-bearing rather than narrative,
fix the explicit keyed-write family

\[
\mathcal{A}_{\mathrm{ts}}^{\mathrm{write}} = \{a^{\mathrm{ts}}_{k,p,\phi} : k \in K,\ p \in P,\ \phi \in \Phi\}
\]

and the write admissibility domain

\[
\operatorname{Adm}_{\mathrm{ts}}^{\mathrm{write}} = \mathcal{M}_{\mathrm{ts}} \times \mathcal{A}_{\mathrm{ts}}^{\mathrm{write}}.
\]

The corresponding update operator is

\[
U_{\mathrm{ts}}\bigl((s,\varphi), a^{\mathrm{ts}}_{k,p,\phi}\bigr) = \bigl(s[k \mapsto p], \varphi[k \mapsto \phi]\bigr),
\]

where \(s[k \mapsto p]\) and \(\varphi[k \mapsto \phi]\) denote keyed
overwrite. Let \(\mathcal{A}_{\mathrm{ts}}\) be any ambient action set
containing this write family. For the theorem-bearing coarse fragment,
no further admissibility or update clauses are fixed here. Sections 8.4
and 8.6 use only \(\mathcal{A}_{\mathrm{ts}}^{\mathrm{write}}\)
operationally, and Section 8.5 uses additional response labels only at
the abstract-signature level.

Let

\[
\mathcal{L}^{\mathrm{write}} = \{\operatorname{write}(k,p,\phi) : k \in K,\ p \in P,\ \phi \in \Phi\} \subseteq \mathcal{A}^{\ast}
\]

be the shared visible-action skeleton used by the response layer on this
worked fragment, and define the coarse concretization map

\[
\iota_{\mathrm{ts}} : \mathcal{L}^{\mathrm{write}} \to \mathcal{A}_{\mathrm{ts}}^{\mathrm{write}}, \qquad \iota_{\mathrm{ts}}(\operatorname{write}(k,p,\phi)) = a^{\mathrm{ts}}_{k,p,\phi}.
\]

Extend \(\iota_{\mathrm{ts}}\) by \(\bot\) to a map

\[
\iota_{\mathrm{ts}}^{\bot} : \mathcal{L}^{\mathrm{write}} \cup \{\bot\} \to \mathcal{A}_{\mathrm{ts}}^{\mathrm{write}} \cup \{\bot\},
\]

with \(\iota_{\mathrm{ts}}^{\bot}(\bot) = \bot\). For the visible-write
family used in this worked fragment, require the already fixed coarse
abstraction map \(\alpha\) to satisfy
\(\alpha \circ \iota_{\mathrm{ts}} = \mathrm{id}_{\mathcal{L}^{\mathrm{write}}}\).

For each agent \(i\), fix a finite relevant key set \(K_i \subseteq K\),
a summary space \(C_i\), and a summary map
\(c_i : \mathcal{M}_{\mathrm{ts}} \to C_i\). Define the
response-relevant feature surface

\[
W_i^{\mathrm{ts}} = \bigl(K_i \rightharpoonup P\bigr) \times \Phi^{K_i} \times C_i
\]

and the feature map

\[
\Pi_i^{\mathrm{ts}}(x) = \bigl(s|_{K_i}, \varphi|_{K_i}, c_i(x)\bigr) \in W_i^{\mathrm{ts}}.
\]

Take the observation space for agent \(i\) on this fragment to be
\(\mathcal{V}_i^{\mathrm{ts}} = W_i^{\mathrm{ts}}\), and set

\[
O_i^{\mathrm{ts}} = \Pi_i^{\mathrm{ts}}.
\]

Choose maps

\[
\widehat{P}_i : \mathcal{V}_i^{\mathrm{ts}} \to \{0,1\}, \qquad \widehat{R}_i^{\mathrm{vis}} : \mathcal{V}_i^{\mathrm{ts}} \to \mathcal{L}^{\mathrm{write}} \cup \{\bot\},
\]

and define

\[
P_i^{\mathrm{ts}} = \widehat{P}_i, \qquad R_i^{\mathrm{ts}} = \iota_{\mathrm{ts}}^{\bot} \circ \widehat{R}_i^{\mathrm{vis}}.
\]

Thus

\[
\Gamma_i^{\mathrm{ts}} = R_i^{\mathrm{ts}} \circ O_i^{\mathrm{ts}} = \iota_{\mathrm{ts}}^{\bot} \circ \widehat{R}_i^{\mathrm{vis}} \circ \Pi_i^{\mathrm{ts}},
\]

so the tuple-space medium has an explicit observation operator,
activation predicate, and response map, with the response layer
factoring through the same response-relevant query surface and the same
visible-action skeleton used later by the refined medium.

\hypertarget{timestamped-virtual-stigmergic-dataspace}{%
\subsection{8.2 Timestamped virtual-stigmergic
dataspace}\label{timestamped-virtual-stigmergic-dataspace}}

Let \(T\) be a totally ordered timestamp domain and \(\mathcal{S}\) a
finite set of propagation or consistency markers. A virtual-stigmergic
state is a triple

\[
\widetilde{x} = (s,\varphi,\theta)
\]

where \((s,\varphi) \in \mathcal{M}_{\mathrm{ts}}\) and
\(\theta : K \rightharpoonup (T \times \mathcal{S})\) records the
metadata carried with the visible payload store. Let

\[
\pi : \mathcal{M}_{\mathrm{vs}} \to \mathcal{M}_{\mathrm{ts}}, \qquad \pi(s,\varphi,\theta) = (s,\varphi)
\]

be the forgetting map.

For the theorem-bearing refined fragment, fix the matched virtual write
family

\[
\mathcal{A}_{\mathrm{vs}}^{\mathrm{write}} = \{\widetilde{a}^{\mathrm{vs}}_{k,p,\phi} : k \in K,\ p \in P,\ \phi \in \Phi\}
\]

with admissibility domain

\[
\operatorname{Adm}_{\mathrm{vs}}^{\mathrm{write}} = \mathcal{M}_{\mathrm{vs}} \times \mathcal{A}_{\mathrm{vs}}^{\mathrm{write}}.
\]

For each triple \((k,p,\phi)\), let

\[
\mu_{k,p,\phi} : \mathcal{M}_{\mathrm{vs}} \to T \times \mathcal{S}
\]

be a metadata-assignment map. Define the virtual write update by

\[
\widetilde{U}_{\mathrm{vs}}\bigl((s,\varphi,\theta), \widetilde{a}^{\mathrm{vs}}_{k,p,\phi}\bigr)
=
\bigl(s[k \mapsto p], \varphi[k \mapsto \phi], \theta[k \mapsto \mu_{k,p,\phi}(s,\varphi,\theta)]\bigr).
\]

Let \(\mathcal{A}_{\mathrm{vs}}\) be any ambient action set containing
this write family. Define the refined concretization map

\[
\iota_{\mathrm{vs}} : \mathcal{L}^{\mathrm{write}} \to \mathcal{A}_{\mathrm{vs}}^{\mathrm{write}}, \qquad \iota_{\mathrm{vs}}(\operatorname{write}(k,p,\phi)) = \widetilde{a}^{\mathrm{vs}}_{k,p,\phi}.
\]

Extend \(\iota_{\mathrm{vs}}\) by \(\bot\) to a map

\[
\iota_{\mathrm{vs}}^{\bot} : \mathcal{L}^{\mathrm{write}} \cup \{\bot\} \to \mathcal{A}_{\mathrm{vs}}^{\mathrm{write}} \cup \{\bot\},
\]

with \(\iota_{\mathrm{vs}}^{\bot}(\bot) = \bot\). For the visible-write
family used in this worked fragment, require the already fixed refined
abstraction map \(\widetilde{\alpha}\) to satisfy
\(\widetilde{\alpha} \circ \iota_{\mathrm{vs}} = \mathrm{id}_{\mathcal{L}^{\mathrm{write}}}\).
In the theorem-bearing refined fragment below, the only explicit update
clauses are these writes together with the guarded freshness family
introduced in Section 8.4A. Section 8.5 may extend
\(\mathcal{A}_{\mathrm{vs}}\) further by formal labels used only for
abstract-signature separation, not for extra dynamic update clauses. No
further refined admissibility or update clauses are fixed here.

In the timestamp-insensitive regime, keep the same response-relevant
surface as in the tuple-space case and take the observation space for
agent \(i\) to be \(\mathcal{V}_i^{\mathrm{vs}} = W_i^{\mathrm{ts}}\).
Set

\[
O_i^{\mathrm{vs}} = \Pi_i^{\mathrm{ts}} \circ \pi,
\]

\[
P_i^{\mathrm{vs}} = \widehat{P}_i, \qquad R_i^{\mathrm{vs}} = \iota_{\mathrm{vs}}^{\bot} \circ \widehat{R}_i^{\mathrm{vis}}.
\]

Then

\[
\Gamma_i^{\mathrm{vs}} = R_i^{\mathrm{vs}} \circ O_i^{\mathrm{vs}} = \iota_{\mathrm{vs}}^{\bot} \circ \widehat{R}_i^{\mathrm{vis}} \circ \Pi_i^{\mathrm{ts}} \circ \pi,
\]

and because the coarse and refined abstraction maps both collapse their
concrete writes back to the same visible-action skeleton
\(\mathcal{L}^{\mathrm{write}}\), one has for every agent \(i\),

\[
\widetilde{\Sigma}_i^{\mathrm{vs}}(\widetilde{x}) = \Sigma_i^{\mathrm{ts}}(\pi(\widetilde{x})).
\]

Thus the virtual medium is a metadata refinement of the tuple-space
medium in the sense of Theorem 7.8, now with the relevant operators
written explicitly and with the response layer typed through a shared
visible-action skeleton rather than by identifying the two concrete
action sets.

\hypertarget{fiber-alignment-and-quotient-adequacy}{%
\subsection{8.3 Fiber alignment and quotient
adequacy}\label{fiber-alignment-and-quotient-adequacy}}

Write
\(\Sigma^{\mathrm{ts}}(x) = (\Sigma_i^{\mathrm{ts}}(x))_{i \in \mathcal{I}}\)
and
\(\widetilde{\Sigma}^{\mathrm{vs}}(\widetilde{x}) = (\widetilde{\Sigma}_i^{\mathrm{vs}}(\widetilde{x}))_{i \in \mathcal{I}}\).

In the timestamp-insensitive regime, every fiber of \(\pi\) is
response-aligned. Indeed, if

\[
\pi(\widetilde{x}) = x,
\]

then the virtual and tuple-space signatures satisfy

\[
\widetilde{\Sigma}^{\mathrm{vs}}(\widetilde{x}) = \Sigma^{\mathrm{ts}}(x).
\]

Therefore Theorem 7.8 applies with
\(\rho = \mathrm{id}_{\mathcal{M}_{\mathrm{ts}}}\) and
\(\widetilde{\rho} = \pi\). The coarse payload-phase medium
\(\mathcal{M}_{\mathrm{ts}}\) itself is response-adequate for the pair,
and in particular the timestamped virtual-stigmergic dataspace is
response-equivalent to its tuple-space quotient whenever timestamps and
propagation markers are invisible to abstract enabled response.

This is the positive half of the boundary result: the quotient works
exactly because every refined fiber carries the same abstract response
signature as its coarse image.

\hypertarget{matched-writes-and-one-step-preservation}{%
\subsection{8.4 Matched writes and one-step
preservation}\label{matched-writes-and-one-step-preservation}}

This subsection gives the concrete witness for Theorem 7.9. Consider
keyed write actions

\[
a^{\mathrm{ts}}_{k,p,\phi} \in \mathcal{A}_{\mathrm{ts}},
\qquad
\widetilde{a}^{\mathrm{vs}}_{k,p,\phi} \in \mathcal{A}_{\mathrm{vs}},
\]

where the tuple-space action writes payload \(p\) and phase tag \(\phi\)
at key \(k\), while the virtual action performs the same visible write
and additionally assigns fresh metadata at \(k\). By construction, both
concrete actions abstract to the same visible write label
\(\operatorname{write}(k,p,\phi)\), and let

\[
B_{\mathrm{write}} = \{(a^{\mathrm{ts}}_{k,p,\phi}, \widetilde{a}^{\mathrm{vs}}_{k,p,\phi}) : k \in K,\ p \in P,\ \phi \in \Phi\}.
\]

For every
\((a^{\mathrm{ts}}_{k,p,\phi}, \widetilde{a}^{\mathrm{vs}}_{k,p,\phi}) \in B_{\mathrm{write}}\)
and every virtual state \(\widetilde{x} = (s,\varphi,\theta)\), the
explicit operators of Sections 8.1--8.2 give

\[
\begin{aligned}
\pi\bigl(\widetilde{U}_{\mathrm{vs}}(\widetilde{x},\widetilde{a}^{\mathrm{vs}}_{k,p,\phi})\bigr)
&= \pi\bigl(s[k \mapsto p], \varphi[k \mapsto \phi], \theta[k \mapsto \mu_{k,p,\phi}(s,\varphi,\theta)]\bigr) \\
&= \bigl(s[k \mapsto p], \varphi[k \mapsto \phi]\bigr) \\
&= U_{\mathrm{ts}}\bigl(\pi(\widetilde{x}), a^{\mathrm{ts}}_{k,p,\phi}\bigr).
\end{aligned}
\]

Thus \(B_{\mathrm{write}}\) is an explicit instance of the exogenous
matched-action construction required by Theorem 7.9: its members are
precisely the write pairs that agree on visible key, payload, and phase.
The quotient comparison also has a one-step dynamic part. For this
concrete \(B_{\mathrm{write}}\), matched writes commute with the
forgetting map, and, provided every fiber of \(\pi\) is
response-aligned, preserve response-equivalence at the successor state.
Without that fiber-alignment hypothesis only the projected-successor
identity survives, not successor-state response equivalence.

\hypertarget{a-guarded-freshness-actions-and-failure-of-matched-representation}{%
\subsection{8.4A Guarded freshness actions and failure of matched
representation}\label{a-guarded-freshness-actions-and-failure-of-matched-representation}}

This subsection gives the concrete witness for Theorem 7.9A. The
positive relation \(B_{\mathrm{write}}\) of Section 8.4 depends on
unconditional writes. As soon as the visible effect of a virtual action
is guarded by hidden freshness metadata, the matched-action picture can
collapse.

Enlarge the virtual action set, if needed, by the guarded family

\[
\mathcal{A}_{\mathrm{vs}}^{\mathrm{guard}} = \{\widetilde{a}^{\mathrm{vs,guard}}_{k,p,\phi,\tau} : k \in K,\ p \in P,\ \phi \in \Phi,\ \tau \in T\}.
\]

For the enlarged explicit fragment, set

\[
\operatorname{Adm}_{\mathrm{vs}}^{\mathrm{write+guard}}
=
\mathcal{M}_{\mathrm{vs}} \times \bigl(\mathcal{A}_{\mathrm{vs}}^{\mathrm{write}} \cup \mathcal{A}_{\mathrm{vs}}^{\mathrm{guard}}\bigr).
\]

On this action family, write

\[
W_{k,p,\phi}(s,\varphi,\theta) = \bigl(s[k \mapsto p], \varphi[k \mapsto \phi], \theta[k \mapsto \mu_{k,p,\phi}(s,\varphi,\theta)]\bigr).
\]

\[
t_{\mathrm{time}}(u,\sigma) = u, \qquad G_{k,\tau}(\theta) = 1 \iff \theta(k) \text{ exists and } t_{\mathrm{time}}(\theta(k)) \ge \tau.
\]

All guarded state-action pairs used below lie in
\(\operatorname{Adm}_{\mathrm{vs}}^{\mathrm{write+guard}}\). Extend the
virtual update operator on this domain by

\[
\widetilde{U}_{\mathrm{vs}}\bigl((s,\varphi,\theta), \widetilde{a}^{\mathrm{vs,guard}}_{k,p,\phi,\tau}\bigr)
=
\begin{cases}
W_{k,p,\phi}(s,\varphi,\theta), & \text{if } G_{k,\tau}(\theta) = 1, \\
(s,\varphi,\theta), & \text{otherwise.}
\end{cases}
\]

Choose a visible tuple-space state \((s,\varphi)\) and parameters
\((k,p,\phi,\tau)\) such that the visible write is nontrivial,

\[
\bigl(s[k \mapsto p], \varphi[k \mapsto \phi]\bigr) \neq (s,\varphi).
\]

Assume the timestamp domain \(T\) contains at least one element strictly
below \(\tau\) and at least one element at or above \(\tau\), and fix
one such pair. Now choose refined states

\[
\widetilde{x}_{<} = (s,\varphi,\theta_{<}), \qquad \widetilde{x}_{\ge} = (s,\varphi,\theta_{\ge})
\]

with the same visible tuple-space projection, with
\(t_{\mathrm{time}}(\theta_{<}(k))\) strictly below the threshold
\(\tau\), and with \(t_{\mathrm{time}}(\theta_{\ge}(k))\) at or above
it. Then

\[
\pi(\widetilde{x}_{<}) = \pi(\widetilde{x}_{\ge}) = (s,\varphi),
\]

but the guarded action produces different projected successors:

\[
\pi\bigl(\widetilde{U}_{\mathrm{vs}}(\widetilde{x}_{<}, \widetilde{a}^{\mathrm{vs,guard}}_{k,p,\phi,\tau})\bigr) = (s,\varphi),
\]

\[
\pi\bigl(\widetilde{U}_{\mathrm{vs}}(\widetilde{x}_{\ge}, \widetilde{a}^{\mathrm{vs,guard}}_{k,p,\phi,\tau})\bigr) = \bigl(s[k \mapsto p], \varphi[k \mapsto \phi]\bigr).
\]

Because these coarse successors are distinct, Theorem 7.9A applies.
Therefore no coarse action \(a \in \mathcal{A}_{\mathrm{ts}}\) can match
\(\widetilde{a}^{\mathrm{vs,guard}}_{k,p,\phi,\tau}\) uniformly on the
fiber over \((s,\varphi)\), and no matched-action relation extending the
quotient comparison can include this guarded action class.

This is the one-step obstruction counterpart to the matched-write result
of Section 8.4. The quotient can still represent unconditional writes,
but freshness-conditional writes already fail at the level of one-step
matched evolution. The same mechanism scales: multi-threshold or
multi-key freshness guards create multiple response-distinguishing
classes over one coarse state, which is the higher-multiplicity case
used in Section 8.5 and Theorem 7.10A.

\hypertarget{canonical-coarsest-repair-and-forced-lower-bounds-for-freshness-sensitive-fibers}{%
\subsection{8.5 Canonical coarsest repair and forced lower bounds for
freshness-sensitive
fibers}\label{canonical-coarsest-repair-and-forced-lower-bounds-for-freshness-sensitive-fibers}}

\textbf{Static signature-level witness, not a dynamic enlargement.} This
subsection supplies the concrete witness for Theorem 7.10 together with
the lower-bound witness for Theorem 7.10A. It changes only the response
layer for one distinguished agent and is a static signature-level
witness for Theorems 7.8, 7.10, and 7.10A. The executable write/guard
fragment used in Sections 8.4 and 8.6 is not enlarged here; the commit
labels introduced below are formal comparison labels only.

The quotient breaks as soon as visible response depends on the relative
freshness of several hidden keys. Fix an integer \(r \ge 2\), distinct
keys \(k_1,\dots,k_r \in K\), and a distinguished agent \(j\). Take the
common abstract comparison alphabet \(\mathcal{A}^{\ast}\) to contain
\(\mathcal{L}^{\mathrm{write}}\) together with distinct symbols
\(\operatorname{commit}(k_1), \dots, \operatorname{commit}(k_r)\). Keep
the timestamp-insensitive operators from Section 8.2 for every agent
\(i \neq j\). Using the timestamp projection \(t_{\mathrm{time}}\)
introduced in Section 8.4A, define the partial component

\[
\widehat{t}(\theta,k) =
\begin{cases}
t_{\mathrm{time}}(\theta(k)), & \text{if } \theta(k) \text{ is defined,} \\
\mathrm{undef}, & \text{otherwise,}
\end{cases}
\]

and write

\[
\Xi_j(s,\varphi,\theta) = \bigl(\Pi_j^{\mathrm{ts}}(s,\varphi),\; \widehat{t}(\theta,k_1), \dots, \widehat{t}(\theta,k_r)\bigr).
\]

The observation space for agent \(j\) under this enriched regime is
\(\mathcal{V}_j^{\mathrm{vs,max}} = W_j^{\mathrm{ts}} \times (T \cup \{\mathrm{undef}\})^r\),
where \(T\) is the timestamp domain and \(\mathrm{undef}\) denotes an
undefined timestamp entry. Because each component
\(\widehat{t}(\theta,k_q)\) lies in \(T \cup \{\mathrm{undef}\}\) by
construction, the map \(\Xi_j\) lands in
\(\mathcal{V}_j^{\mathrm{vs,max}}\).

Write \(q^{\ast}(\theta) = q\) when all projected timestamps
\(\widehat{t}(\theta,k_1),\dots,\widehat{t}(\theta,k_r)\) are defined
(equivalently, each \(\theta(k_q)\) is defined) and \(k_q\) is the
unique freshest key among \(k_1,\dots,k_r\) under the total order on
\(T\), and leave \(q^{\ast}(\theta)\) undefined otherwise. Replace the
distinguished observation and response pair by

\[
O_j^{\mathrm{vs,max}}(s,\varphi,\theta) = \Xi_j(s,\varphi,\theta),
\]

\[
P_j^{\mathrm{vs,max}}(\Xi_j(s,\varphi,\theta)) = 1,
\]

and

\[
R_j^{\mathrm{vs,max}}(\Xi_j(s,\varphi,\theta))
=
\begin{cases}
\operatorname{commit}(k_{q^{\ast}(\theta)}), & \text{if } q^{\ast}(\theta) \text{ is defined,} \\
\bot, & \text{otherwise.}
\end{cases}
\]

When \(q^{\ast}(\theta)\) is undefined, because some timestamps are
missing or because no key is uniquely freshest, the \(\bot\) response is
a deliberate modeling choice: it represents abstention under ambiguity,
not a typing gap. Concretely, the pair
\((P_j^{\mathrm{vs,max}}, R_j^{\mathrm{vs,max}}) = (1, \bot)\) encodes
``agent \(j\) is active but abstains,'' and contributes no element to
the enabled-response set \(E_{\widetilde{\mathfrak{S}}}^{\ast}\), while
the alternative configuration \((0, \bot)\) would encode ``agent \(j\)
is disabled.'' Only the active-abstention regime is used here.

so that

\[
\Gamma_j^{\mathrm{vs,max}} = R_j^{\mathrm{vs,max}} \circ O_j^{\mathrm{vs,max}}.
\]

Assume moreover that the refined abstraction map keeps these
distinguished commit labels separate at the abstract level, so that

\[
\widetilde{\alpha}(\operatorname{commit}(k_q)) \neq \widetilde{\alpha}(\operatorname{commit}(k_{q'}))
\qquad
\text{for } q \neq q'.
\]

Extend \(\mathcal{A}_{\mathrm{vs}}\) further, if needed, by auxiliary
formal labels
\(\{\operatorname{commit}(k_q) : q = 1,\dots,r\} \subseteq \mathcal{A}_{\mathrm{vs}}\),
and on this auxiliary family take \(\widetilde{\alpha}\) to be the
evident inclusion into \(\mathcal{A}^{\ast}\). These commit labels are
included only as formal comparison labels in the ambient refined action
alphabet. No pair involving one of them is placed in the admissibility
domain of the worked update fragment, so they are excluded from the
executable update domain and are used only in the static signature-level
witness of this subsection, not as additional dynamic update clauses.
The coarse abstraction map \(\alpha\) need not hit any
\(\operatorname{commit}(k_q)\); the repair theorems use a common
abstract codomain, not a common image of \(\alpha\) and
\(\widetilde{\alpha}\).

Choose a visible tuple-space state \((s,\varphi)\) and refined states

\[
\widetilde{x}^{(q)} = (s,\varphi,\theta^{(q)})
\qquad
\text{for } q = 1,\dots,r,
\]

such that for each \(q\), all keys \(k_1,\dots,k_r\) are defined in
\(\theta^{(q)}\) and \(k_q\) is the unique freshest key among them. Such
choices exist when the timestamp domain \(T\) contains at least \(r\)
distinct elements: total order on \(T\) then lets us choose \(r\)
distinct timestamp values and assign them so that \(k_q\) is freshest in
\(\theta^{(q)}\). We add this cardinality hypothesis \(|T| \ge r\) here.
Then

\[
\pi(\widetilde{x}^{(1)}) = \cdots = \pi(\widetilde{x}^{(r)}) = (s,\varphi),
\]

while the distinguished agent satisfies

\[
\Gamma_j^{\mathrm{vs,max}}(\widetilde{x}^{(q)}) = \operatorname{commit}(k_q)
\]

for every \(q\). All other agents retain the timestamp-insensitive
signatures from Section 8.2, and the displayed abstraction-side
distinctness assumption ensures that the distinguished abstract
responses remain pairwise distinct after applying
\(\widetilde{\alpha}\). Therefore the full refined signature tuples

\[
\widetilde{\Sigma}^{\mathrm{vs}}(\widetilde{x}^{(1)}), \dots, \widetilde{\Sigma}^{\mathrm{vs}}(\widetilde{x}^{(r)})
\]

are pairwise distinct. Consequently,

\[
|F_{\pi}(s,\varphi)| \ge r,
\]

and the fiber over \((s,\varphi)\) is not response-aligned. Theorem 7.8
therefore fails exactly because hidden freshness data now determines
visible response.

Theorem 7.10 instead supplies the canonical coarsest response-adequate
repair

\[
\mathcal{C}_{\min}^{\mathrm{max}} = \{((s,\varphi),\sigma) : (s,\varphi) \in \mathcal{M}_{\mathrm{ts}},\; \sigma \in F_{\pi}(s,\varphi) \cup \{\Sigma^{\mathrm{ts}}(s,\varphi)\}\},
\]

with projection and decoders

\[
p_{\min}^{\mathrm{max}}(((s,\varphi),\sigma)) = (s,\varphi), \qquad \Lambda_i^{\min,\mathrm{max}}(((s,\varphi),\sigma)) = \sigma_i,
\]

and representations

\[
\rho_{\min}^{\mathrm{ts}}(s,\varphi) = ((s,\varphi),\Sigma^{\mathrm{ts}}(s,\varphi)),
\]

\[
\widetilde{\rho}_{\min}^{\mathrm{vs}}(s,\varphi,\theta) = ((s,\varphi),\widetilde{\Sigma}^{\mathrm{vs}}(s,\varphi,\theta)).
\]

The new point is that the repair size is now forced. For every
response-adequate repair

\[
(\mathcal{C},\rho,\widetilde{\rho},p,\{\Lambda_i\}_{i \in \mathcal{I}})
\]

of this freshness-sensitive quotient, Theorem 7.10A gives

\[
|p^{-1}(s,\varphi)| \ge |F_{\pi}(s,\varphi) \cup \{\Sigma^{\mathrm{ts}}(s,\varphi)\}| \ge r.
\]

And for the canonical repair itself,

\[
|(p_{\min}^{\mathrm{max}})^{-1}(s,\varphi)| = |F_{\pi}(s,\varphi) \cup \{\Sigma^{\mathrm{ts}}(s,\varphi)\}|.
\]

The repair therefore separates more than one binary freshness condition.
Its fiber size grows with the number of distinct freshness-sensitive
response classes carried by the quotient fiber, and the canonical repair
is exactly as small as the theory allows.

\hypertarget{finite-matched-serial-traces}{%
\subsection{8.6 Finite matched serial
traces}\label{finite-matched-serial-traces}}

This subsection records the concrete serial closure of Section 8.4
rather than a separate new construction. Theorem 7.11 now gives a
pathwise witness for the timestamp-insensitive regime. Fix a virtual
execution fragment consisting of matched write actions,

\[
\widetilde{x}_0 \xrightarrow{\widetilde{a}^{\mathrm{vs}}_{k_0,p_0,\phi_0}} \widetilde{x}_1 \xrightarrow{\widetilde{a}^{\mathrm{vs}}_{k_1,p_1,\phi_1}} \cdots \xrightarrow{\widetilde{a}^{\mathrm{vs}}_{k_{n-1},p_{n-1},\phi_{n-1}}} \widetilde{x}_n,
\]

and for each step choose the matched tuple-space action

\[
a^{\mathrm{ts}}_{k_t,p_t,\phi_t} \in \mathcal{A}_{\mathrm{ts}}
\]

so that

\[
\bigl(a^{\mathrm{ts}}_{k_t,p_t,\phi_t}, \widetilde{a}^{\mathrm{vs}}_{k_t,p_t,\phi_t}\bigr) \in B_{\mathrm{write}}
\]

for every \(t\). Set

\[
x_0 = \pi(\widetilde{x}_0), \qquad x_{t+1} = U_{\mathrm{ts}}(x_t,a^{\mathrm{ts}}_{k_t,p_t,\phi_t}).
\]

Because each step lies in \(B_{\mathrm{write}}\) and Section 8.4
verifies the hypotheses of Theorem 7.9 for that relation, Theorem 7.11
gives

\[
\pi(\widetilde{x}_t) = x_t \quad \text{for all } t = 0,\dots,n.
\]

Hence

\[
\widetilde{\Sigma}^{\mathrm{vs}}(\widetilde{x}_t) = \Sigma^{\mathrm{ts}}(x_t)
\]

for every step, and therefore the abstract enabled-response traces agree
pointwise along the full fragment. In other words, a finite serial write
trace in the timestamped virtual medium projects to a unique tuple-space
trace with the same abstract response behavior at every state visited.
This is the concrete witness of Theorem 7.11 for the dataspace pair.

\hypertarget{additional-downstream-specializations}{%
\subsection{8.7 Additional downstream
specializations}\label{additional-downstream-specializations}}

The same comparison pattern extends beyond the worked pair. Linked-data
media and ledger-state media can often be compared through a coarser
query or status surface when graph metadata, consensus metadata, or
block-level details are not visible to enabled response
\citep{linkeddata2021, stigld2022, gurcan2022pow}. Conversely, when
those fields alter enabled response, Theorem 7.10 identifies the
canonical coarsest adequate repair as the coarse quotient refined by the
corresponding response-fiber classes. We do not develop those additional
cases fully here; they are recorded as downstream realizations of the
same abstract scheme.

\hypertarget{limits}{%
\section{9. Limits}\label{limits}}

The manuscript develops a systems-theory comparison layer rather than a
finished theory of everything stigmergic. Several limitations are
explicit.

First, only one cross-medium pair is worked out in detail. The
tuple-space and virtual-stigmergy comparison is intended as a
theorem-bearing exemplar, not as a complete classification of stigmergic
media.

Second, the comparison theory now reaches finite serial matched traces
but no further. We prove a quotient adequacy criterion for aligned
fibers, a canonical coarsest-repair theorem together with fiberwise
lower bounds for freshness-visible refinements, one-step update
preservation for matched writes, a one-step non-liftability obstruction
for freshness-conditional actions, and a finite matched-trace lift for
explicitly matched serial executions, but do not yet provide a branching
simulation theory, cost-sensitive equivalence, or robustness analysis.

Third, only a minimal serial execution model is fixed, intended for
turn-based or otherwise single-threaded shared-medium interactions. The
treatment therefore avoids making scheduler-heavy liveness claims
central to the theorem spine and does not attempt a general theory of
concurrent interleavings; the first nontrivial extension would be
fixed-order interleavings or scheduler-parameterized serializations.

Fourth, the operator taxonomy of Section 5 is descriptive rather than
theorem-bearing. It organizes interpretation and examples and should be
read as such; it is explicitly not a formal classification theorem and
supplies no separation or completeness claims.

Finally, the worked dataspace pair uses a simplified keyed shared
medium. That level of abstraction is deliberate: the point is to isolate
the response-relevant comparison structure before pursuing richer
medium-specific semantics.

\hypertarget{conclusion}{%
\section{10. Conclusion}\label{conclusion}}

Stigmergy already has strong conceptual foundations and multiple formal
literatures
\citep{heylighen2016stigmergy1, dipple2014general, denicola2020virtual, stigld2022, boldini2024control}.
What has remained comparatively underdeveloped is a medium-agnostic
comparison layer that treats shared-medium coordination itself as a
state-transition class open to admissibility analysis and
representation.

The strongest contribution here is accordingly more specific than a
universal theory of stigmergy. We identify abstract enabled-response
signatures as the right comparison object, show that a coarse quotient
is response-adequate if and only if its fibers are response-aligned,
construct the canonical coarsest response-adequate repair when
freshness-visible metadata breaks the quotient, derive fiberwise lower
bounds on any response-adequate repair of a failing quotient, establish
one-step preservation for matched writes whose visible effects commute
with the quotient, exhibit a one-step non-liftability obstruction for
freshness-conditional actions, and record the finite serial closure of
the matched one-step result. The explicit tuple-space and
virtual-stigmergy instantiation confirms that the abstract layer yields
a concrete cross-medium adequacy-and-repair cycle rather than a purely
interpretive catalogue of substrates.

That is the sense in which the manuscript aims to contribute to the
mathematical study of stigmergic coordination: not by replacing the
existing semantic and application-specific literatures, but by supplying
a sharper comparison language, one worked positive/negative comparison
cycle with both a dynamic action-level obstruction and a repair-level
obstruction component, and one explicit finite serial closure result
that future media-specific realizations can test, extend, or refute.

\hypertarget{acknowledgments-and-disclosure}{%
\section{Acknowledgments and
Disclosure}\label{acknowledgments-and-disclosure}}

The author used AI writing assistance during drafting and editing. All
mathematical definitions, theorem statements, proofs, and final claims
are the author's responsibility.

\bibliographystyle{plainnat}
\bibliography{references}

\end{document}